\documentclass[USenglish,oneside,twocolumn]{article}

\pdfoutput=1

\usepackage[utf8]{inputenc}
\usepackage[big]{dgruyter_NEW}

\usepackage{mempatch}
\usepackage{graphicx}
\usepackage{url}
\usepackage{multirow}

\usepackage[labelfont=bf,justification=justified]{caption} 
\usepackage[caption=false]{subfig} 

\usepackage[usenames,dvipsnames]{color}
\graphicspath{{figures/}{graphs/}}
\DeclareGraphicsExtensions{.pdf,.jpeg,.png}

\usepackage{wrapfig}
\usepackage{xspace}
\usepackage{algorithm}
\usepackage{algpseudocode}
\usepackage{varwidth}
\DeclareCaptionType{copyrightbox}
\usepackage{enumitem}

\usepackage{xcolor}
 
\cclogo{\includegraphics{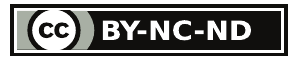}}
  
\newcommand{\etal}{{\em et al.}}

\renewcommand{\S}[1]{Section~\ref{#1}}
\renewcommand{\paragraph}[1]{\noindent{\textbf{#1:}}}

\newcommand{\pluseq}{\mathrel{{+}{=}}}
\newcommand{\minuseq}{\mathrel{{-}{=}}}

\newcommand{\rt}[1]{\textcolor{red}{#1}}

\begin{document}

  \author*[1]{John Geddes}

  \author[2]{Mike Schliep}

  \author[3]{Nicholas Hopper}

  \affil[1]{University of Minnesota, E-mail: geddes@cs.umn.edu}

  \affil[2]{University of Minnesota, E-mail: schliep@cs.umn.edu}

  \affil[3]{University of Minnesota, E-mail: hopper@cs.umn.edu}

\title{Anarchy in Tor: Performance Cost of Decentralization}

\runningtitle{Anarchy in Tor: Performance Cost of Decentralization}


  \begin{abstract}
{
    Like many routing protocols, the Tor anonymity network has
    decentralized path selection, in clients locally and independently
    choose paths.  As a result, network resources may
    be left idle, leaving the system in a suboptimal state.  This is referred to
    as the price of anarchy, where agents acting in their own self
    interest can
    make poor decisions when viewed in a global context.  In this paper we
    explore the cost of anarchy in Tor by examining the potential performance
    increases that can be gained by centrally optimizing circuit and relay selection
    using global knowledge.  In experiments with both offline and
    online algorithms, we show that centrally coordinated
    clients can achieve up to 75\% higher bandwidth compared to traditional
    Tor.  Drawing on these findings, we design and evaluate a decentralized version of our
    online algorithm, in which relays locally distribute information enabling clients to
    make smarter decisions locally and perform downloads 10-60\% faster.
    Finally, we perform a privacy analysis of the decentralized algorithm
    against a passive and active adversary trying to reduce anonymity
    of clients and increase their view of the Tor network.  We conclude
    that this decentralized algorithm does not enable new attacks,
    while providing significantly higher performance.
}
\end{abstract}



\maketitle

\section{Introduction} \label{sec:introduction}

Tor \cite{tor-design} is one of the most popular and widely used low-latency
anonymity systems, with 2 million daily clients supported by a network of over
6000 volunteer relays.  Tor clients tunnel traffic through a sequence of 3
relays called a \emph{circuit}.  Traffic sent over the circuit is encrypted
so that no single relay can learn the identity of both the client and
destination.  With a limited amount of resources available from the volunteer
relays, it is important to distribute traffic to utilize
these resources as efficiently as possible.  Doing so ensures high performance to latency sensitive applications
and attracts potential clients to use the network, increasing anonymity for all
users.

Similar to ordinary routing, clients make decisions locally based on their
view of the network.  In Tor, directory authorities are centralized servers
that keep track of all relays and their information.  Clients download this
information, stored in consensus files and server descriptors, every few hours
from the directory authorities.  With this information clients create 10
circuits to have available, selecting relays for the circuits at random
weighted by the relays bandwidth. When the client starts using Tor to
send traffic, it then selects the circuit in best standing to use for the next
10 minutes.  Since it can take the directory authority up to 24 hours to
properly update relay information and distribute it to clients, much
research has focused on how clients can make better decisions, such as
considering latency \cite{oakland2012-lastor} or relay congestion
\cite{congestion-tor12} when selecting relays and circuits.

The main issue with leaving network decisions to clients is that local
incentives can lead clients to make decisions that \emph{they} believe will
maximize their performance, when actually these decisions result in a global
sub-optimal state with overall client performance suffering.  In these
situations it is common to compare what happens when decisions are made with
global knowledge.  This can be achieved by centrally processing requests as
they are made, or performing decisions offline when the entire set of requests
is known a priori.  The performance gap that results from using decentralized
compared to centralized decision making is called the \textit{price of
anarchy}.  

While Tor clients are not necessarily selfish, they are making local decisions
without context of the state of the global network.  In this paper we
analyze the price of anarchy in the Tor network.  We develop both offline
and online centralized algorithms that optimize relay and circuit selection decisions made
by clients.  Using the global view of all active circuits in the Tor network,
the algorithms are able to more intelligently select which circuits clients
should use for each download they perform, resulting in significant performance
improvements to clients.  While these are not necessarily protocols that should be run in
practice, it allows us to bound the actual price of anarchy in Tor.  This
demonstrates the potential improvements that better resource allocation in Tor
could achieve.

In this paper we make the following major contributions:
\begin{itemize}
\item Using a modified client model, we create a genetic algorithm to
  compute the optimal offline circuit selection for a given sequence
  of download requests.  Since the algorithm is given access to the
  full sequence of requests before processing, this allows us to
  establish the optimal performance for a given network and load.
  This serves as a baseline for the competitive analysis of other algorithms.

    \item In addition, we develop an online centralized algorithm that processes
        download requests serially.  Based on a delay-weighted capacity routing
        algorithm the centralized circuit selection tries to avoid low
        bandwidth bottlenecks when assigning circuits to requests.
    \item Using techniques from the centralized algorithm, we develop a
        decentralized algorithm that results in similar resource allocations,
        with relays releasing a small amount of information to allow clients to
        make smarter circuit selection choices.
    \item We perform privacy analysis on the decentralized circuit selection
        algorithm that considers both information that can be learned from a
        passive adversary, and examines how an active adversary could attempt
        to abuse the algorithm to increase their view of the network.
    \item All algorithms are analyzed in a large scale simulated environment
        using the Shadow simulator \cite{jansen2012shadow}.  This allows us to
        precisely compare the effects and performance gains achieved across the
        different circuit selection algorithms.
\end{itemize}

\section{Background} \label{sec:background}

In this section we discuss some of the Tor architecture, basics on how the
clients operate and send data through the network, and related work
on increasing performance in Tor.

\subsection{Tor Architecture}

When a client first joins the Tor network, it downloads a list of relays with
their relevant information from a directory authority.  The client then
creates 10 \emph{circuits} through a guard, middle, and exit relay.  The relays
used for each circuit are selected at random weighted by bandwidth.
For each TCP connection the client wishes to make, Tor creates a \emph{stream}
which is then assigned to a circuit; each circuit will typically handle
multiple streams concurrently. The Tor client will find a viable circuit that
can be used for the stream, which will be then be used for the next 10 minutes
or until the circuit becomes unusable.

Internal to the overlay network, relay pairs establish a single TLS connection
for \emph{all} communication between them.  
To send traffic through a circuit, data is packed into 512-byte onion-crypted
cells using secret keys negotiated with each relay during the circuit building
process.  Once a relay has received a cell, it peels off its layer of
encryption, finds the next relay on the circuit to send to, and places it on a
circuit queue where it waits to be sent.  After a cell has traversed the entire
circuit, the exit recovers the initial data sent by the client and is forwarded
to the end destination.


\subsection{Related Work} 
One of the major keys to increasing anonymity for Tor
users is ensure a large anonymity set, that is, a large user base.  To do so Tor
needs to offer low latency to the clients; bad performance in the form of slow
web browsing can lead to fewer users using the system overall.  To this end,
there has been a plethora of research looking to address ways to increase
performance in Tor.  These roughly fall into 4 areas: scheduling, selection,
transports, and incentives.

\paragraph{Scheduling} Internally Tor is constantly making scheduling decisions
on what to send and how much should be processed.  These decisions happen on
all levels, between streams, circuits, and connections.  On the circuit level,
much work has been done in an attempt to classify latency sensitive circuits,
either prioritizing them \cite{ccs10-scheduling,ccs2012-classification} or
outright throttling noisy ones \cite{throttling-sec12}.  With regards to
deciding \emph{how much} should be sent on a circuit, there has been work
comparing Tor's end-to-end window algorithm with an ATM-style link-based
algorithm \cite{pets2011-defenestrator}.  Additional work \cite{jansen14-kist}
has been done on scheduling across circuits from \emph{all} connections,
limiting how much is written to a connection at a time so Tor has more control
over what gets sent, opposed to the kernel.


\paragraph{Selection} When creating circuits, Tor selects from relays at random
weighted by their bandwidth.  Determining the bandwidth is a non-trivial issue,
and much work \cite{snader2011improving,snader2009eigenspeed} has been done
looking at a range of methods, from using self-reported values, central nodes
making direct measurements, and peer-to-peer methods relying on existing
relays. There also has been a lot of research in improving the relay selection
\cite{sherr2009scalable,sherr2010a3,oakland2012-lastor,congestion-tor12,
ndss13-relay-selection,ccs2014-mators,herbert2014optimising} for circuit
creation.  These range from incorporating latency and congestion measurements,
using a virtual coordinate system, to adjusting the weighting strategy, all in
an attempt improve overall client performance.


\paragraph{Transport} One of the noted performance issues in Tor is the
fact that the single TLS connection between relays can cause unnecessary
blocking, where circuits could keep sending data but TCP mechanisms prevent it
\cite{dingledine2009performance}. Reardon \cite{reardon-thesis} attempted to
address this by implementing TCP-over-DTLS allowing a connection to be
dedicated to each circuit.  In a similar vein, there has been numerous
work \cite{wpes12-torchestra,ccs2013-pctcp,wpes14-imux} looking into increasing
the number of TCP connections between relays and scheduling circuits between
them in an attempt to avoid unneccessary blocking.  Nowlan \etal 
introduce uTCP and uTLS\cite{nowlan2012fitting,nowlan2013reducing} , which
allows for out-of-order delivery in Tor so it can process cells from circuits
that are not blocking.


\paragraph{Incentives} While the previous lines of research involved improving
efficiency in how Tor handles traffic, another set looked at potential ways to
incentive clients to also contribute bandwidth as a relay, increasing the
overall network resources available to clients.  This was first explored by
Ngan, Dingledine, and Wallach \cite{incentives-fc10}, where they would
prioritize traffic from relays providing high quality service in the Tor
network.  Jansen, Hopper, and Kim \cite{ccs10-braids} extend this idea,
allowing relays to earn credits which can be redeemed for higher prioritized traffic.
Building on this, Jansen, Johnson, and Syverson \cite{ndss13-lira} introduce a
more lightweight solution that allows for the same kind of prioritization of
traffic without as much overhead.

\section{Experimental Setup} \label{sec:experiments}

To maximize our understanding of performance in the Tor network, we want to run
large scale experiments that accurately represent the actual operations of Tor
clients and relays.  To this end we use Shadow
\cite{shadowdev,jansen2012shadow}, a discrete event network simulator with the
capability to run actual Tor code in a simulated network environment.  Shadow
allows us to setup a large scale network configuration of clients, relays, and
servers, which can all be simulated on a single machine.  This lets us run
experiments privately without operating on the actual Tor network, avoiding
potential privacy concerns of dealing with real users.  Additionally, it lets
us have a global view and precise control over every aspect of the network,
giving us insights and capabilities into Tor that would be near impossible to
achieve in a live network.  Most importantly, Shadow performs deterministic
runs, allowing for reproducible results and letting us isolate exactly what we
want to test for performance effects.

Our experiments are configured to use Shadow v1.9.2 and Tor v0.2.5.10.  We use
the large network configuration deployed with Shadow which consists of 500
relays, 1350 web clients, 150 bulk clients, 300 performance clients, and 500
servers.  The default client model in Shadow
\cite{jansen2012shadow,jansen2012methodically,jansen14-kist} has clients
download a file of a specific size, and after the download is finished it
chooses how long to pause until it starts the next download.  Web clients
download 320 KB files and will randomly pause between 1 and 60,0000
milliseconds until it starts the next download.  Bulk clients download 5 MB
files with no break in between downloads.  The performance clients are split
into three groups, downloading 50 KB, 1 MB, and 5 MB files, then pausing 1
minute between each download it performs.  Along with the default client model
we also consider a new type of client model called the fixed download model.
Instead of clients having a single start time and inter-download pause times,
they have a \textit{list} of downloads with each download having a unique start
time and end time.  To create a fixed download experiment we extract the start
and stop times for all client downloads from a default client run and use those
as the base for the fixed download experiment.

For measuring performance between experiments we examine different metrics
depending on which client model is being used.  For the default client model we
look at the time to first byte and download times of web and bulk clients along
with the total read bandwidth across all clients in the network.  When using
the new fixed download model, download times lose meaning as they have fixed
start and end times.  The major metric that will change between experiments is
the amount of data being pushed through the Tor network, so total client read
bandwidth is the only metric we consider when using the fixed download model.

\begin{figure*}[t]
    \centering
    \includegraphics[width=1.0\textwidth]{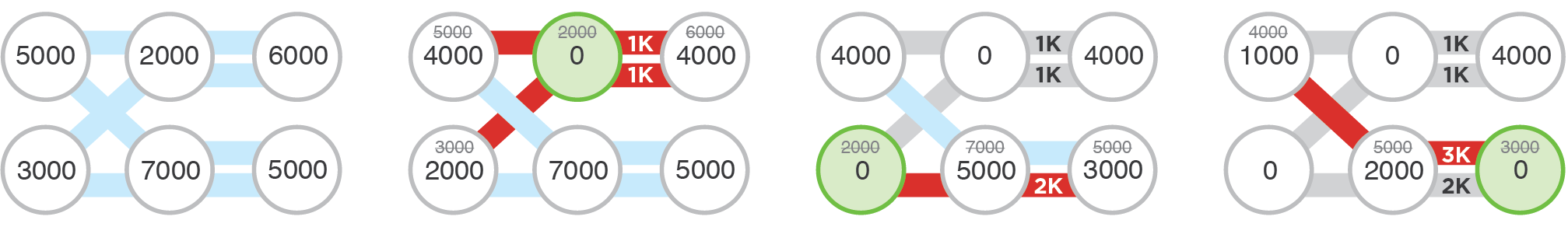}
    \caption{Example of circuit bandwidth estimate algorithm.  Each step the relay with
    the lowest bandwidth is selected, its bandwidth evenly distributed among
any remaining circuit, with each relay on that circuit having their bandwidth
decremented by the circuit bandwidth.}
    \label{fig:bandwidth-calc}
\end{figure*}

\section{Price of Anarchy} \label{sec:anarchy}

The price of anarchy \cite{roughgarden2005selfish} refers to situations where
decentralized decision making can lead a system into a sub-optimal
configuration due to the selfish nature of agents participating in the system.
In networking this problem is specifically referred to as selfish routing
\cite{roughgarden2002bad,roughgarden2002selfish}, where users select routes
based on a local optimization criteria (latency, bandwidth, etc.) resulting in a
sub-optimal global solution due to potential conflicting interests.  While
clients in Tor are not necessarily \emph{selfish}, as they select relays at
random weighted by their bandwidth, they are \emph{oblivious} with respect to
relay and network conditions.  In that sense, localized decision making with
respect to relays and circuits has the potential to lead to a sub-optimal result,
where network resources are left idle which could be used to increase client
performance.

To fully explore the price of anarchy in Tor we want to examine just how much
of a performance increase can be achieved if we have some centralized authority making
decisions rather than the clients themselves.  For this we consider both an
\textit{offline} algorithm, which is able to see the entire set of download
requests when operating, and an \textit{online} algorithm which needs to process inputs
serially and is unaware of future requests.  In this section we detail how each
of these algorithms work and look at any potential performance benefits.

\subsection{Offline Algorithm} \label{ssec:offline}

An offline algorithm needs access to the entire input set while operating, in
this case that means knowing the times of when each client is active and
performing a download.  So instead of using the default client model in Shadow,
we use the fixed download model which gives us a list of all download start and
stop times in the experiment.  From the experimental setup we extract
all downloads $d_i$ with their corresponding start and end times $(s_i,
e_i)$, along with the list of relays $r_j$ and their bandwidth $bw_j$.  These
parameters are then passed into the offline algorithm which returns a mapping
$d_i \rightarrow c_i(r_{i_1},r_{i_2},r_{i_3})$ of downloads and a circuit
consisting of three relays which should be used by the client performing the
download.  Note that we still use the constraints from Tor that all relays in
the circuit must be unique and that $r_{i_3}$ must be an exit relay, but we do
not force the algorithm to select $r_{i_1}$ to be a guard.  This is done since
it is impossible to use a non-exit as an exit relay, however there is no
mechanism to enforce the guard in a circuit actually has the guard flag set.

When using the fixed download client model the offline circuit selection
problem starts to strongly resemble job scheduling
\cite{graham1979optimization}.  For example, we have a set of downloads (jobs)
that have to be processed on three relays (machines).  Each download has a
start time (release date), and can either have an end time (due date) or a file
of specific size to download (number of operations).  The relays can process
downloads simultaneously (jobs can be preempted) and the goal is to download
each file as fast as possible (minimize complete time or lateness).  There are
some complications in how exactly to define the machine environment for job
scheduling, since the relays are heterogeneous and the fact that the amount of
processing done on a relay is determined by the bottleneck in the circuit and
not necessarily the bandwidth capacity of the relay.  We would definitely need
an algorithm which handled job splitting across machines, as we need 3 relays
processing a download, and these problems seem to be often NP-hard.

Due to these complications we instead develop a genetic algorithm in order to
compute a lower-bound of an optimal offline solution.  Our genetic algorithm
will have a population of solutions with a single solution consisting of a
complete mapping of downloads to circuits.  We can breed two solutions by iterating
through each download and randomly choosing a circuit from either parent.
However, the most important part of a genetic algorithm is a fitness function
allowing us to score each solution.  Since the metric we are most concerned
with in the fixed download client model is the amount of bandwidth being pushed
through the network, we want a method for estimating the total amount of
network bandwidth used given a mapping of downloads to circuits.

In order to calculate the total bandwidth across an entire set of downloads, we
first need a method for calculating the circuit bandwidth across a set of
active circuits at a single point of time. To do this we make two observations:
(1) bottleneck relays will determine the bandwidth of the circuit, and (2) that
a relay will split its bandwidth equally among all circuits it is the
bottleneck on.  The first observation comes from the end-to-end window based
congestion control algorithm used in Tor, ensuring clients and exit relays only
send as much as a circuit can handle.  The second observation isn't trivially
true, since with the priority circuit scheduler some circuits will be sending
more than others on a small enough time interval.  In aggregate though all
circuits will be given roughly the same amount of bandwidth. 

\begin{algorithm}[t]
    \caption{Estimate bandwidth of active circuits}
    \label{alg:calccircbw}
    \begin{algorithmic}[1]
        \Function{CalcCircuitBW}{$activeCircuits$}
            \State $activeRelays \gets GetRelays(activeCircuits)$
            \While{\textbf{not }$activeCircuits.empty()$}
                \State $r \gets GetBottleneckRelay(activeRelays)$
                \State $circuits \gets GetCircuits(activeCircuits, r)$
                \State $circuitBW \gets r.bw / circuits.len()$
                \For{$c \in circuits$}
                    \State $c.bw \gets circuitBW$
                    \For{$circRelay \in c.relays$}
                        \State $circRelay.bw \minuseq c.bw$
                        \If{$circRelay.bw = 0$}
                            \State $activeRelays.remove(circRelay)$
                        \EndIf
                    \EndFor
                    \State $activeCircuits.remove(c)$
                \EndFor 
            \EndWhile
        \EndFunction
    \end{algorithmic}
\end{algorithm}

The pseudocode for calculating the circuit bandwidth across a set of active
circuits is shown in Algorithm~\ref{alg:calccircbw} and works as follows.
First the algorithm identifies the relay $r$ with the lowest bandwidth per
circuit (line 4).  We know this relay is the bottleneck on all circuits that
$r$ appears on, as by definition every other relay on the circuit will have a
higher available circuit bandwidth.  Next the algorithm iterates through each
circuit $c_i$ that $r$ appears on and assigns the circuit bandwidth as $r$'s
per circuit bandwidth (lines 5-8).  In addition, for each $c_i$ the algorithm
also iterates over each relay $r_{i_j}$ on circuit $c_i$, decrementing the
bandwidth of $r_{i_j}$ by the circuit bandwidth assigned to circuit $c_i$ (line
10).  While iterating over the relays, if any of them runs out of available
bandwidth, the relay is removed from the list of active relays (lines 11-13).
After we have iterated over the relays in circuit $c_i$, it is removed from the
list of active circuits (line 15).  Note that during this process relay $r$
will always end up with a bandwidth of 0, and no circuits remaining in the list
of active circuits will contain $r$.  Furthermore, any \textit{other} relay
that reached a bandwidth value of 0 will also not be contained on any circuit
remaining in the active circuit list \footnote{The proof for this is shown in
Appendix\ref{sec:bwalgproof}}.  This means that we can never have a situation
where a circuit has a relay that is no longer active with a bandwidth of 0,
resulting in a circuit bandwidth of 0.  The algorithm repeats this process,
extracting the bottleneck relay and updating relay and circuit bandwidths,
until no active circuits remain.  An example of this this algorithm is shown in
Figure~\ref{fig:bandwidth-calc}, highlighting which relay is selected and how
the circuit bandwidth is assigned and relay bandwidth updated.

\begin{algorithm}[t]
    \caption{Compute total bandwidth for downloads}
    \label{alg:calctotalbw}
    \begin{algorithmic}[1]
        \Function{CalcTotalBW}{$downloads$}
            \State $start,end \gets GetTimeInterval(downloads)$
            \State $bandwidth \gets 0$
            \State $time \gets start$
            \While{$time \leq end$}
                \State $circuits \gets GetActiveCircs(downloads, time)$
                \State $CalcCircuitBW(circuits)$
                \For{$circuit \in circuits$}
                    \State $bandwidth \pluseq circuit.bw$
                \EndFor
                \State $time \gets time + tick$
            \EndWhile
            \State \textbf{return} $bandwidth$
        \EndFunction
    \end{algorithmic}
\end{algorithm}

With an algorithm to compute the bandwidth across all active circuits at a
single point in time, we can now compute the overall bandwidth consumption
across an entire set of downloads with assigned circuits, which is outlined in
Algorithm~\ref{alg:calctotalbw}. First the algorithm retrieves the earliest
start time and latest end time across all downloads (line 2).  With this we
then iterate over every ``tick'' between the start and end time (lines 5-12),
calculating the bandwidth across all active circuits during that tick (lines
6-7).  Then for each active circuit we simply add the circuit bandwidth to a
total bandwidth counter (lines 8-10).  This gives us a total bandwidth value
for a circuit to download mapping that can be used as the fitness score for
solutions in the genetic algorithm.  With this fitness score we can now run the
genetic algorithm across populations of circuit to download mappings.  The main
parameters for the genetic algorithm are the breed percentile $b$, indicating
the top $b\%$ of solutions to pick from when breeding, the elite top $e\%$ of
solutions that get copied over into the next generation, and the mutation
probability $m$. For our purposes a mutation can occur when we are selecting a
circuit for a single download, and we simply select a relay in the circuit to
be replaced when another randomly chosen relay.  \footnote{Our experiments used
parameters $b=0.2$ and $e=0.1$.  Mutations were only allowed during some of the
experiments and was set at $m=0.01$.}.

\subsection{Online Algorithm} \label{ssec:online}

The offline genetic algorithm has access to the entire set of downloads when
making circuit selection decisions.  We want to see how it compares to an
online algorithm, one that has to make circuit selection decisions serially.
This way it has global knowledge of which active downloads are using which
circuits, but has no idea how long downloads will last and does not know when
future downloads start.  
While at first glance the online problem might seem similar to traditional
routing, there are some problems using routing algorithm with respect to
circuit selection in Tor.  Typically routing algorithms depict the network as a
graph in order to run efficient algorithms in order to extract which paths to
select.  Since relays, not links, are the resource constrained entities we
can invert the graph, with each relay consisting of and in and out vertex and
the edge between them representing the bandwidth of the relay.  Additionally,
since in Tor there is a link between every relay, for each relay we add an edge
with infinite bandwidth connecting its out vertex with every other relays in
vertex.  But now if we want to run any traditional algorithms (e.g. max flow),
there is not way to impose the requirement that all circuits must contain
exactly three hops.  In order to do this we would have to create a graph of all
potential circuits, where each relay would have a separate vertex for when it
can be a guard, middle, and exit relay.  But not only does this explode our
state space, there is now no way to make sure that when a relay is used as a
guard, the available bandwidth on the middle and exit vertex decreases
correspondingly.

So while we cannot use traditional routing algorithm for online circuit
selection, we can borrow some techniques from routing algorithms that share
similar goals.  Specifically, for our online algorithm we borrow techniques
used in the delay-weighted capacity (DWC) routing
algorithm\cite{yang2001quality}.  The goal of the DWC algorithm is similar to
what we wish to accomplish for circuit selection, that is to find the path with
the highest potential capacity.  There are quite a few differences in the
original DWC algorithm and what we need to accomplish, mainly that we are
\textit{given} a set of circuits to choose from and do not need to extract
paths from a graph.  The key insight we can use from the DWC algorithm is
bottleneck identification and weight assignment.  The algorithm first extracts
a set of least delay paths, then for each path it identifies the bottleneck
link, incrementing the links weight by the inverse bandwidth of the
link. Then the DWC algorithm simply selects the path with the lowest sum of
weights across all links in the path.  By doing this it avoids the most
critical links, those that are the bottlenecks providing the lowest amount of
bandwidth.

To adopt the DWC routing algorithm for an online circuit selection algorithm,
we need a method for identifying which \textit{relays} are bottlenecks on
active circuits.  Algorithm~\ref{alg:calccircbw}, which estimates the bandwidth
for all active circuits, can very easily be adapted for this purpose.  Note
that when we get the relay with the lowest per circuit bandwidth (line 4), we
know that this relay is the bottleneck on each active circuit it appears on.
So when we iterate through those circuits (line 7) and assign them the per
circuit bandwidth value (line 8) we can \textit{also} update the weight for the
relay.  By simply adding 
\begin{eqnarray*} 
    r.weight \gets r.weight + \frac{1}{circuitBW} 
\end{eqnarray*} 
after the circuit bandwidth assignment (between lines 8 and 9).  Now when the
function is done each circuit will have a bandwidth \emph{and} each relay will
have a corresponding DWC weight.  So to run the DWC circuit selection
algorithm, we first iterate through the downloads \textit{in-order} based on
the download start times.  For each new download we get the set of current active
downloads and their assigned circuits.  We then run the updated algorithm that
calculates active circuit bandwidth \textit{and} relay DWC weights.  Then, for
each circuit we are considering for the new download, we calculate the circuit
weight as the sum of relay weights in the circuit.  Then the new download is
assigned the circuit with the lowest calculated weight.  If there are multiple
circuits with the lowest weight we select the circuit with the highest
available bandwidth, defined as the minimum across all relay available
bandwidth values.

\begin{figure}[t!]
    \centering
    \includegraphics[width=0.45\textwidth]{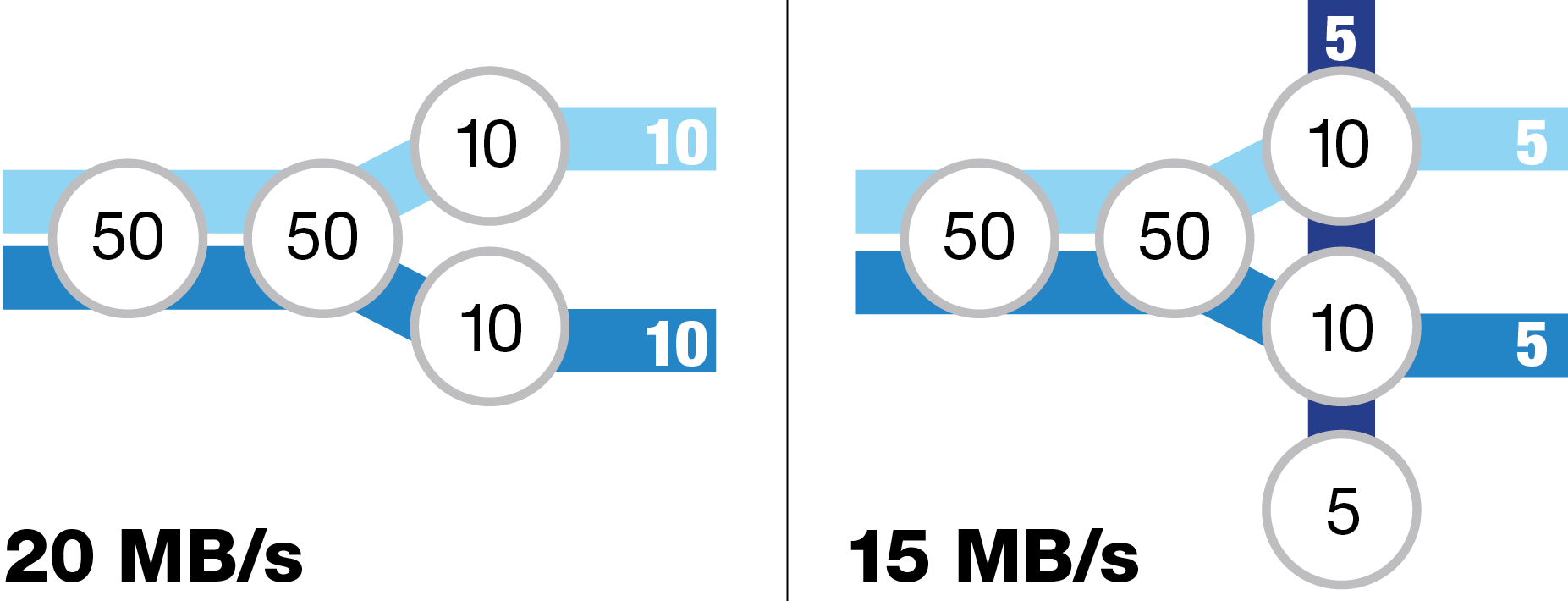}
    \caption{On the left, each circuit can achieve 10 MBps throughput, resulting
    in a total of 20 MBps being delivered to the clients. On the right, both
    original circuits must share bandwidth at the exit nodes with a third
    circuit, reducing the throughput for each circuit to 5 MBps and the total
    network throughput to 15 MBps, a drop in network throughput even though
we added a circuit to the network.}
    \label{fig:pruned-example}
\end{figure}

\begin{figure*}[t!]
    \centering
    \subfloat[Offline Genetic Algorithm]{\label{fig:offline-genetic-results}
       {\includegraphics[width=0.33\textwidth]{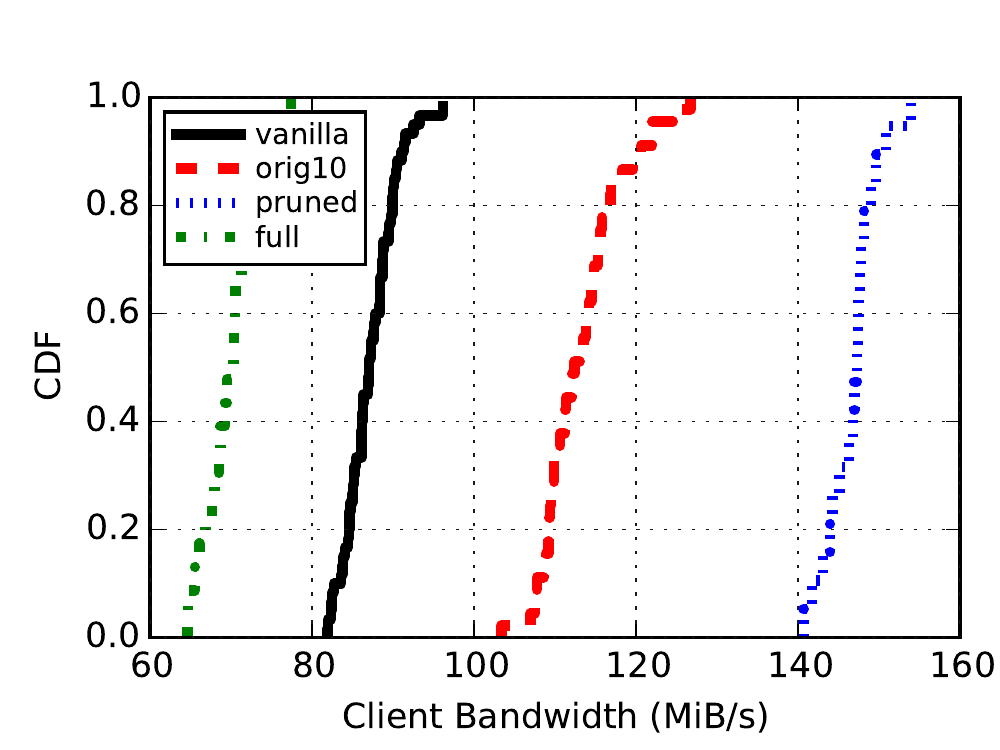}}}
    \subfloat[Online DWC Algorithm]{\label{fig:offline-dwc-results}
       {\includegraphics[width=0.33\textwidth]{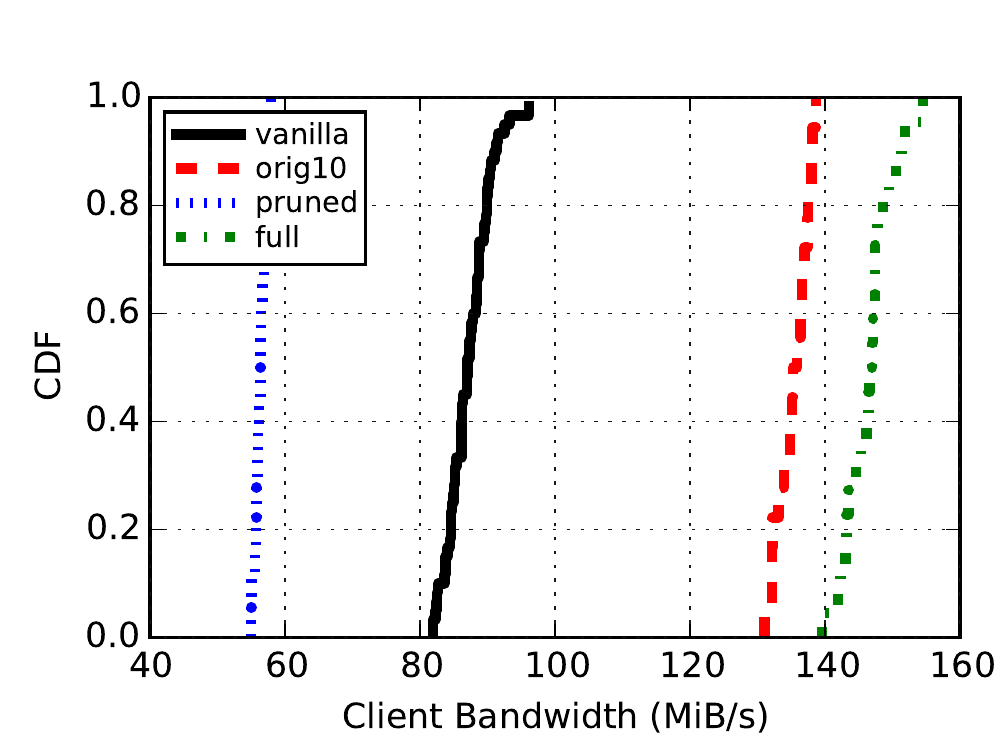}}}
    \subfloat[Relay Capacity]{\label{fig:offline-relay-capacity}
    {\includegraphics[width=0.33\textwidth]{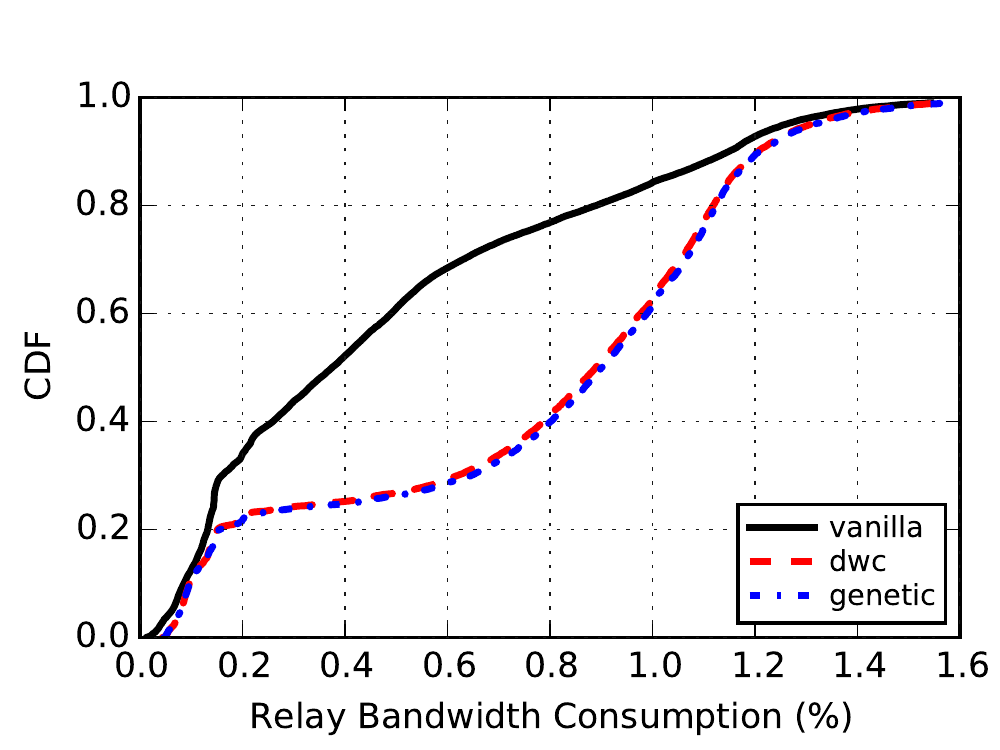}}}
       \caption{Total client bandwidth when using the offline and online
       circuit selection while changing the set of available circuits, along
   with the relay bandwidth capacity of the best results from both algorithms.}
    \label{fig:central}
\end{figure*}

\subsection{Circuit Sets}

In both the offline and online circuit selection algorithms, each download has
a set of potential circuits that can be selected from.  For our experiments we
consider three potential circuit sets.  The first is the original 10 circuits
that were available during the initial vanilla run. Along with extracting the
fixed download start and end times from a completed vanilla experiment, we
\textit{also} record which circuits were available on the client when each
download started.  This allows us to specifically focus on circuit selection,
keeping relay selection identical across all experiments.  The second circuit
set we consider is the full set of all potential valid circuits.  We consider a
circuit $r_1,r_2,r_3$ valid if $r_3$ is an actual exit relay, and if $r_1 \not=
r_2$, $r_2 \not= r_3$, and $r_1 \not= r_3$.  Note we do not require that $r_1$
actually has the guard flag set.  This is because while there are mechanisms
internally in Tor that prevent any use of a non-exit relay being used as an
exit relay, there is nothing stopping a client from using a non-guard relay as
their guard in a circuit.  We also remove any ``duplicate'' circuits that
contain the same relays, just in a different order.  This is done since neither
algorithm considers inter-relay latency, only relay bandwidth, so there is no
need to consider circuits with identical relays.

The final circuit set we consider is the pruned circuit set.  The motivation
behind the pruned set can be seen in Figure~\ref{fig:pruned-example}.  It shows
that we can add an active circuit and actually \textit{reduce} the amount of
bandwidth being pushed through the Tor network. In the example shown it will
always be better to use one of the already active circuits shown on the left
instead of the new circuit seen on the right side of the graphic.  There are
two main ideas behind building the pruned circuit set: (1) when building
circuits always use non-exit relays for the guard and middle if possible and
(2) select relays with the highest bandwidth.  To build a circuit to add to the
pruned set, the algorithm first finds the exit relay with the highest bandwidth.
If none exist, the algorithm can stop as no new circuits can be built.  Once it
has an exit, the algorithm then searches for the non-exit relay with the highest
bandwidth, and this relay is used for the middle relay in the circuit.  If one
is not found it searches the remaining exit relays for the highest bandwidth
relay to use for the middle in the circuit.  If it still cannot find a relay,
the algorithm stops as there are not enough relays left to build another
circuit.  The search process for the middle is replicated for the guard, again
stopping if it cannot find a suitable relay.  Now that the circuit has a guard,
middle, and exit relay, the circuit is added to the pruned circuit set and the
algorithm calculates the circuit bandwidth as the minimum bandwidth across all
relays in the circuit.  Each relay then decrements its relay bandwidth by the
circuit bandwidth.  Any relay that now has a bandwidth of 0 is permanently
removed from the relay list.  This is repeated until the algorithm can no
longer build valid circuits.  \footnote{The pseudocode for this algorithm is
shown in Appendix~\ref{sec:prunealg}.}

\subsection{Results}

In this section we explore the performance of our offline genetic and online
DWC circuit selection algorithms.  To start we first configured and ran a large
scale experiment running vanilla Tor with no modifications, while using the
default client model.  The results of this run were then used to generate an
experiment using the fixed download model, where each download had the same
start and end times that was seen in the vanilla experiment.  We also extracted
the original circuit sets for each download. These download times are then fed
into both algorithms, along with which circuit sets to use,  which then output
a circuit selection for each download.

We ran the offline genetic and online DWC algorithms once for each circuit set,
resulting in mappings of fixed downloads to circuits.  Large scale experiments
were run using each of the outputted circuit selections, and the results were
compared to the original vanilla Tor experimental run.  Recall since we are
using the fixed download client model, the only metric we are concerned with is
network usage, in this case the total client download bandwidth during the
experiment.  Figures~\ref{fig:offline-genetic-results} and
\ref{fig:offline-dwc-results} shows the total client bandwidth achieved with
the genetic and DWC algorithms respectively.  The first thing to note is that
both algorithms were able to produce relay and circuit selections that improved
bandwidth by 72\%, with median client bandwidth jumping from 86 MBps up to 147
MBps.  To see why we can look at relay capacity seen in
Figure~\ref{fig:offline-relay-capacity}.  This looks at the percent of
bandwidth being used on a relay compared to their configured
\verb|BandwidthRate|.  Note that this can go higher than 100\% because Tor
also has a \verb|BandwidthBurstRate| that allows the relay to temporarily send
more than the \verb|BandwidthRate| over short periods of time.  This shows us
that in the best performing DWC and genetic algorithms, half of the time relays
are using 90\% or more of their configured bandwidth, which was only happening
about 20\% of the time in vanilla Tor.  This shows that the algorithms were
able to take advantage of resources that were otherwise left idle in vanilla
Tor.

Interesting while both algorithms were able to achieve near identical maximum
results, these came from different circuit sets.  The genetic algorithm
achieved its best results when using the pruned set while the DWC algorithm
performed best with the full set of valid circuits.  Furthermore, when the
genetic algorithm used the full set and the DWC algorithm used the pruned set,
they both actually performed \textit{worse} when compared to vanilla Tor.  We
suspect that when using the full circuit set, the genetic algorithm gets stuck
in a local maximum that it is unable to escape, and using the pruned circuit
set prevents the algorithm from entering one of these suboptimal local
maximums.  When the algorithms were restricted to the original circuit sets
both algorithms were able to improve performance compared to vanilla Tor.  The
offline genetic algorithm was able to increase client bandwidth to 111 MBps,
while the online DWC algorithm saw performance results close to what was
achieved with the full circuit set, with median client bandwidth at 138 MBps
compared to 147 MBps with the full circuit set.

While the genetic algorithm serves as a lower bound on an optimal offline
solution, given the near identical results seen when using the pruned circuit
set in the genetic algorithm compared with the full circuit set in the DWC
algorithm, we believe the online DWC algorithm is operating fairly close to
an optimal solution, resulting in a much better allocation of resources.  Most
likely to see even more increased performance we would need to either increase
the bandwidth resources available (e.g. add more relays or increase the
bandwidth of existing relays), or add more clients to the network that can
consume any remaining idle resources.


\section{Decentralized Circuit Selection} \label{sec:decentralized}

In the previous section we saw that when making circuit selection in a
centralized fashion, with a global view of the network, we were able to see
large performance increases with resources being used much more efficiently.
In this section we look at a decentralized solution that borrows techniques
from the online DWC algorithm, allowing for clients to make smarter circuit
selections locally.  The main insight from the online DWC algorithm outlined in
\S{ssec:online} is that we want to avoid bottleneck relays, particularly if
they are low-bandwidth bottlenecks.  The obvious advantage from using a
centralized approach is we have a global view of active circuits and can
accurately estimate where the bottlenecks are and how much bandwidth they are
providing per circuit.  In a decentralized setting no one entity has this
global view that can determine who is and is not a bottleneck on each circuit.
So instead we want a method for relays themselves to estimate which circuits
they are a bottleneck on.  With estimates of bottleneck circuits, they can
locally compute their DWC weight and leak it to clients, allowing clients to
make more intelligent circuit selection decisions.

\subsection{Local Weight Computation} \label{ssec:local-weight}

The main challenge in having relays locally compute their DWC weight is having
the relays estimate which circuits they are the bottleneck on.  
To calculate bottleneck circuits, we make three observations: (1) to be a
bottleneck on \emph{any} circuit the relay's bandwidth should be fully
consumed (2) all bottleneck circuits should be sending more cells than
non-bottleneck circuits, and (3) the number of cells being sent on bottleneck
circuits should be fairly uniform.  The last point is due to congestion control
built into Tor, where the sending rate on a circuit should converge to the
capacity of the bottleneck.  So the two main things we need is a method for
calculating the bandwidth of a circuit and a way to classify circuits as
bottleneck or non-bottleneck based on their bandwidth value.

While calculating circuit bandwidth might seem easy, we need to be careful to
treat bulk and web circuits equally.  Bulk clients continuously send data
through the circuit, so those should always be at or close to capacity.  Web
clients are more bursty, sending traffic in short periods and resting for an
unknown amount of time.  In order to assign a circuit a bandwidth we consider
two parameters.  First is the bandwidth \textit{window}, in which we keep track
of all traffic sent on a circuit for the past $w$ seconds.  The other
parameters is the bandwidth \textit{granularity}, a sliding window of $g$ that
we scan over the $w$ second bandwidth window to find the maximum number of
cells transferred during any $g$ period.  That maximum value is then assigned as
the circuit bandwidth. With circuits assigned a bandwidth value we need a way
to cluster the circuits into bottleneck and non-bottleneck groups.  For this we
look at three different clustering methods.

\paragraph{Jenks Natural Breaks}
The Jenks natural breaks \cite{jenks1967data} clustering algorithm attempts to
cluster one-dimensional data into classes that minimizes the in-class variance.
The function takes in a one-dimensional array of data and the $n$ classes that
the data should be clustered into.  It returns the breaks $[b_1, b_2), [b_2,
b_3), ..., [b_n, b_{n+1}]$ and a goodness of variance fit (GVF) $var \in
[0,1]$, where higher variances indicate better fits.  For clustering circuits
we can use the Jenks algorithm in two different ways.  First is to cluster the
data into 2 classes, with circuits in the $[b_1,b_2)$ range classified as
non-bottlenecks and those in $[b_2,b_3]$ classified as bottlenecks.  Second is
we can keep incrementing the number of classes we cluster the circuits into
until the GVF value passes some threshold $\tau$.  Once the threshold is passed
we then classify all circuits in the $[b_n,b_{n+1}]$ range as bottlenecks and
everyone else as non-bottlenecks.

\paragraph{Head/Tail}
The head/tail clustering algorithm \cite{headtail} is useful when the
underlying data has a long tail, which could be useful for bottleneck
identification, as we expect to have a tight clustering around the bottlenecks
with other circuits randomly distributed amongst the lower bandwidth values.
The algorithm first splits the data into two sets, the tail set containing all
values less than the arithmetic mean, and everything greater to or equal to the
mean is put in the head set.  If the percent of values that ended up in the
head set is less than some threshold, the process is repeated using the head
set as the new data set.  Once the threshold is passed the function returns the
very last head set as the head cluster of the data.  The pseudocode for this
algorithm is shown in Algorithm~\ref{alg:headtail}.  For bottleneck
identification we simply pass in the circuit bandwidth data and the head
cluster returned contains all the bottleneck circuits.

\begin{algorithm}[t]
    \caption{Head/Tail clustering algorithm}
    \label{alg:headtail}
    \begin{algorithmic}[1]
        \Function{HeadTail}{$data,threshold$}
            \State $m \gets sum(data) / len(data)$
            \State $head \gets$ \{$d \in data $|$ d \geq m$\}
            \If{$len(head) / len(data) < threshold$}
                \State $head \gets HeadTail(head, threshold)$
            \EndIf
            \State \textbf{return} $head$
        \EndFunction
    \end{algorithmic}
\end{algorithm}

\paragraph{Kernel Density Estimator}
The idea behind the kernel density estimator \cite{kde1,kde2} is we are going
to try and fit a multimodal distribution based on a Gaussian kernel to the
circuits.  Instead of using the bandwidth estimate for each circuit, the
estimator takes as input the entire bandwidth history seen across all circuits,
giving the estimator more data points to build a more accurate density
estimate.  For the kernel bandwidth we initially use the square root of the
mean of all values. Once we have a density we compute the set of local minima
$\{m_1,...,m_n\}$ and classify every circuit with bandwidth above $m_n$ as a
bottleneck.  If the resulting density estimate is unimodal and we do not have
\textit{any} local minima, we repeat this process, halving the kernel bandwidth
until we get a multimodal distribution.

With a method to calculate circuit bandwidth and identify bottleneck circuits,
the relay can calculate their weight in the same way described in
\S{ssec:online}; iterate over each bottleneck circuit and add the inverse of
the bandwidth to the local weight calculation.  This information can be leaked
to clients, allowing them to select the circuit with the lowest weight summed
across all relays in the circuit.

\begin{table*}
    \centering
    \begin{tabular}{ c | c c c c | c c c c }
        Bandwidth Granularity& \multicolumn{4}{ c }{100 ms} & \multicolumn{4}{| c }{1 second}\\
        \hline
        Bandwidth Window & 1s & 2s & 5s & 10s & 1s & 2s & 5s & 10s\\
        \hline
        \hline
        Jenks Two & 1.67 & 1.79 & 3.56 & 8.69  & 1.80 & 2.39 & 5.28 & 13.25\\
       Jenks Best & \rt{1.11} & \rt{1.14} & 1.91 & 3.87  & \rt{1.00} & \rt{1.18} & 2.25 & 4.70\\
      Head/Tail   & \rt{0.74} & \rt{0.92} & 1.66 & 4.36  & \rt{0.90} & \rt{1.13} & 2.45 & 6.14\\
    KernelDensity & \rt{0.81} & \rt{0.82} & \rt{0.85} & \rt{0.93} & \rt{0.79} &
        \rt{0.79} & \rt{0.95} & \rt{1.62}\\
    \end{tabular}
    \caption{The bottleneck clustering methods mean square error across varying
    bandwidth granularity and window parameters, with red values indicating
    scores less than weighted random estimator.}
    \label{tbl:bottlenecks}
\end{table*}

\subsection{Implementation} \label{ssec:local-implementation}

Since the decentralized algorithm needs real time bandwidth information, we
implement the algorithm directly in Tor with clients making circuit selection
decisions as downloads start.  So while we had to use the fixed download client
model in \S{sec:anarchy} in order to precompute circuit selection, for the
decentralized algorithm we can use the default client model where clients have
a single start time and web clients pause randomly between downloads they
perform.  While this means they algorithms need to run in real time, we can now
allows include the time to first byte and file download time in our evaluation
metrics.

To incorporate the decentralized algorithm in Tor we first implemented the
local weight calculation method described in \S{ssec:local-weight}.  As
discussed previously the method has three parameters, bandwidth
granularity $g$, bandwidth window $w$, and which clustering method to use.
For each circuit relays keep a count of how many cells were read and written in
each 100 millisecond interval over the past $w$ seconds.  Then to determine the
circuit bandwidth it iterates through the circuit history to determine the
maximum number of cells sent over a $g$ millisecond period.  The circuit
bandwidth values are then clustered into bottleneck circuits, with the weight
calculated as the sum of inverse circuit bandwidth across all bottlenecks. Now
that a relay can compute their own DWC weight, they need a way to leak this
information to clients.  For this we implement a new cell called a gossip cell.
Each relay records its weight in a gossip cell and appends it on the downstream
queue to the client for each circuit it has.  Since the cell is encrypted no
other relay will be able to modify the weight.  To prevent gossip cells
from causing congestion, relays send the gossips cells on circuits every 5
seconds based on the circuit ID; specifically when $now() \equiv circID \mod
5$.  For clients we modified the \verb|circuit_get_best| function to select
circuits with the lowest weight, using circuit bandwidth as the tie breaker.
While compiling a list of ``acceptable'' circuits that Tor normally selected
from, we additionally keep track of the minimal circuit weight across the
acceptable circuits, and additionally filter out circuits with larger weight
values.  If there are multiple circuits remaining after this, we iterate
through the remaining circuits selecting circuits with the highest circuit
bandwidth, defined as the minimum advertised relay bandwidth across all relays
in the circuit.  If after this step we still have more than one circuit
available we default to Tor's circuit selection among the remaining circuits.

For the centralized DWC algorithm we can no longer precompute circuit selection
since we are using the default client model, so we need a way to run the
centralized algorithm during the experiment as clients start downloads.
To accomplish this we introduce a new central authority node
which clients will need to query for circuit selection decisions.  When a
client starts the central authority connects to the client using the Tor
control protocol.  It disables the client from making circuit selections
decisions locally, and then listens on the \verb|CIRC| and \verb|STREAM|
events.  With the \verb|CIRC| events the central authority know what circuits
are available on every client in the network. The
\verb|STREAM| events then notify the central authority when a client begins a
download, at which time the central authority can select which circuit the
client has that they should use for the new download.  Through the
\verb|STREAM| events the central authority is \textit{also} notified when the
stream has completed so the central authority can remove the circuit assigned
to the stream from the active circuit list.  With this the central authority
obtains a global view of the network with knowledge of every current active
circuit.  So it can use the Algorithm~\ref{alg:calccircbw} discussed in
\S{ssec:online} to compute which relays are bottlenecks on each circuit, update
the weight of relays accordingly, and select the lowest weighted circuit that
clients have available.

\begin{figure*}[t]
    \centering
    \subfloat[Time to First Byte]
    {
        \label{fig:local-central-ttfb}
        {\includegraphics[width=0.23\textwidth]{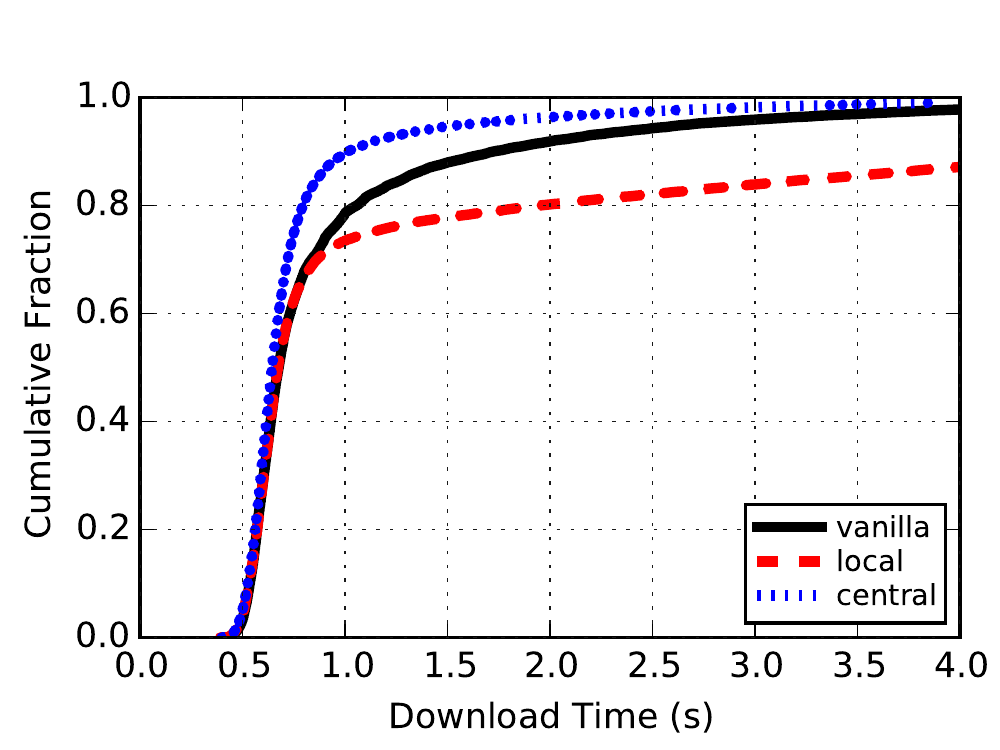}}
    }
    \subfloat[Web Download Times]
    {
        \label{fig:local-central-ttlb-web}
        {\includegraphics[width=0.23\textwidth]{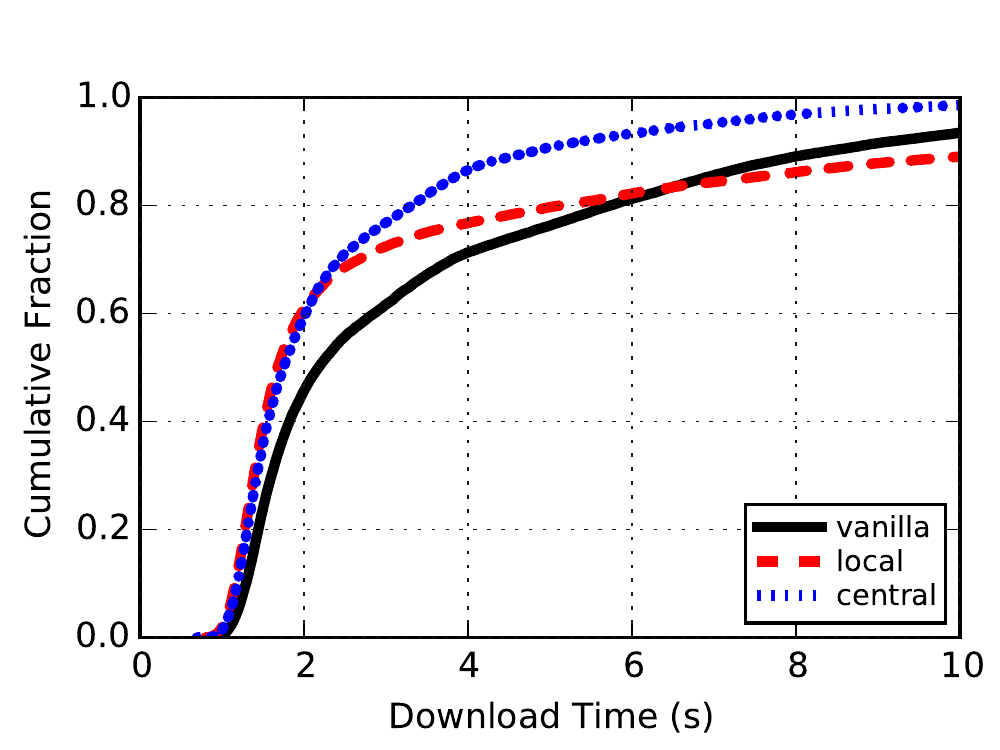}}
    }
    \subfloat[Bulk Download Times]
    {
        \label{fig:local-central-ttlb-bulk}
        {\includegraphics[width=0.23\textwidth]{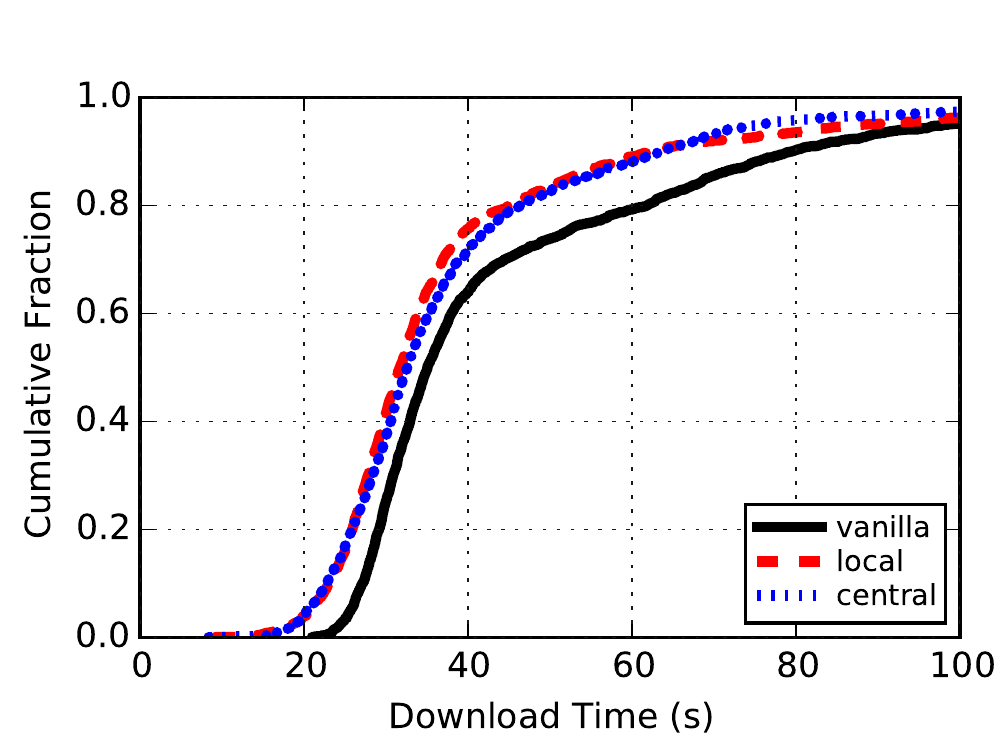}}
    }
    \subfloat[Total Client Bandwidth]
    {
        \label{fig:local-central-client-bw}
        {\includegraphics[width=0.23\textwidth]{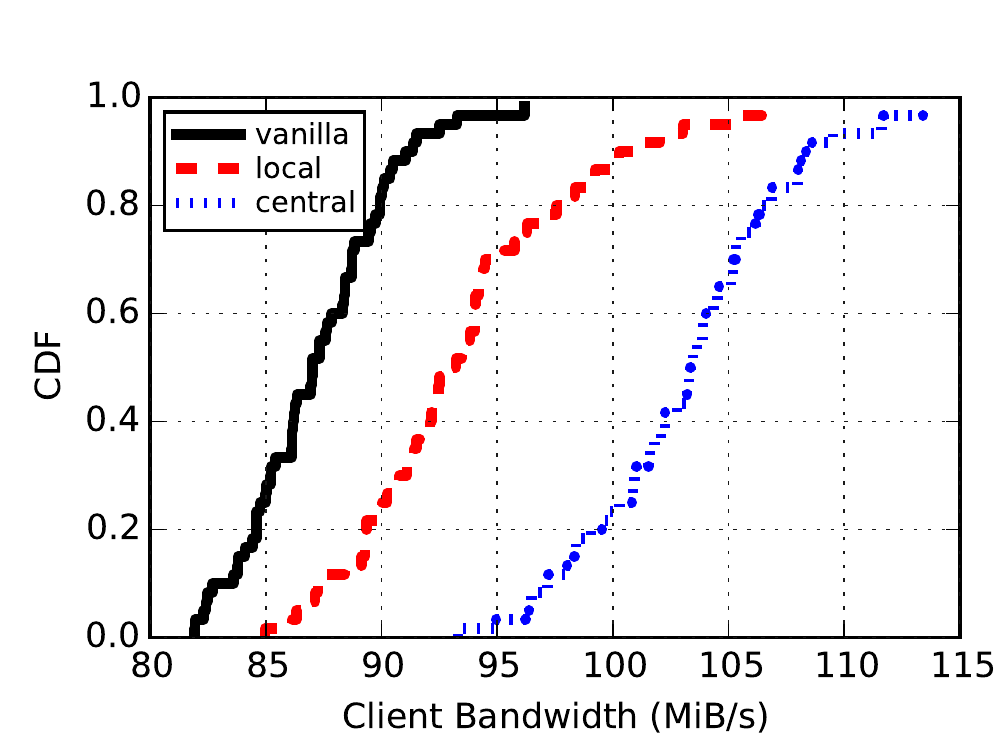}}
    }
    \caption
    {
        Download times and client bandwidth compared across circuit selection
        in vanilla Tor, decentralized, and centralized algorithms.
    }
    \label{fig:local-central-results}
\end{figure*}


\subsection{Results} \label{ssec:local-results}

With the various parameters and clustering methods available to the
decentralized algorithm, the first thing we are interested in is which
parameters results in the most accurate bottleneck estimation.  Since we are
comparing the decentralized algorithm against the centralized DWC algorithm, we
will use the central authorities bottleneck estimation as ground truth.  To
make the comparison we configured an experiment to run with the central
authority making circuit selections.  Every time the central authority selects
a circuit for a client it outputs the bottleneck estimation for every relay.
Additionally, during the run all relays periodically output the entire 10
second bandwidth history for every circuit it is on.  With this information we
can compute which circuits every relay would have classified as a bottleneck
depending on the parameters used.  For comparing estimates, every time the
central authority outputs their bottleneck figures, every relay that outputs
their circuit history within 10 simulated milliseconds will have their local
bottleneck estimate compared to the estimate produced by the central authority.

We are interested in the combination of parameters and clustering methods that
minimizes the mean-squared error between the central authority and local relay
bottleneck estimations. Table~\ref{tbl:bottlenecks} shows the results of the
different clustering methods with bandwidth granularity set at either 100
millisecond or 1 second, and with the bandwidth window varying between 1,2,5
and 10 seconds.  For comparison we created a weighted random estimator that
simply randomly selects from the entire set of estimates produced by the
central authority.  The weighted random estimator produced a mean-squared error
of 1.33, which was better than almost half of all configurations.
For example the Jenks two class clustering method consistently produced results
worse than the weighted random estimator no matter what parameters were used.
Across all clustering methods larger bandwidth windows resulted in higher
mean-squared errors.  This indicates that circuits that were inactive (and thus
excluded from the central authorities calculation) were still being assigned a
high bandwidth value and clustered into the bottleneck group.
Interestingly, while the kernel density estimator produced the best results
\textit{on average}, the overall best parameter configuration was using the
head/tail estimator with bandwidth granularity $g=100ms$ and bandwidth window
$w=1s$.

\begin{figure*}[t]
    \centering
    \subfloat[Time to First Byte]
    {
        \label{fig:local-ncirc-ttfb}
        {\includegraphics[width=0.23\textwidth]{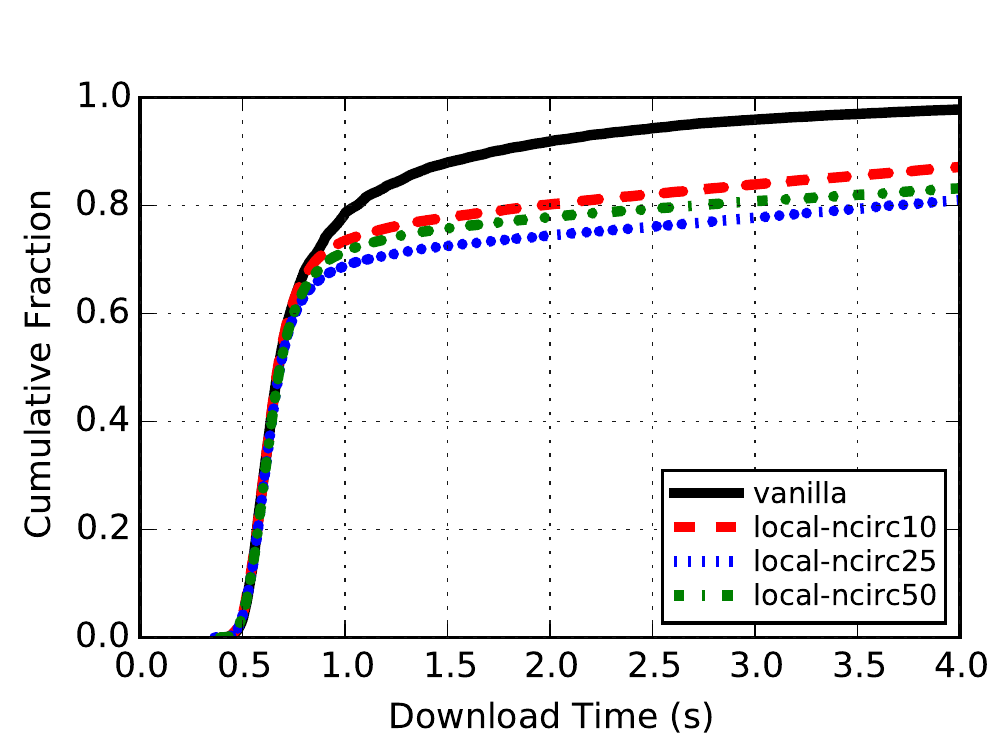}}
    }
    \subfloat[Web Download Times]
    {
        \label{fig:local-ncirc-ttlb-web}
        {\includegraphics[width=0.23\textwidth]{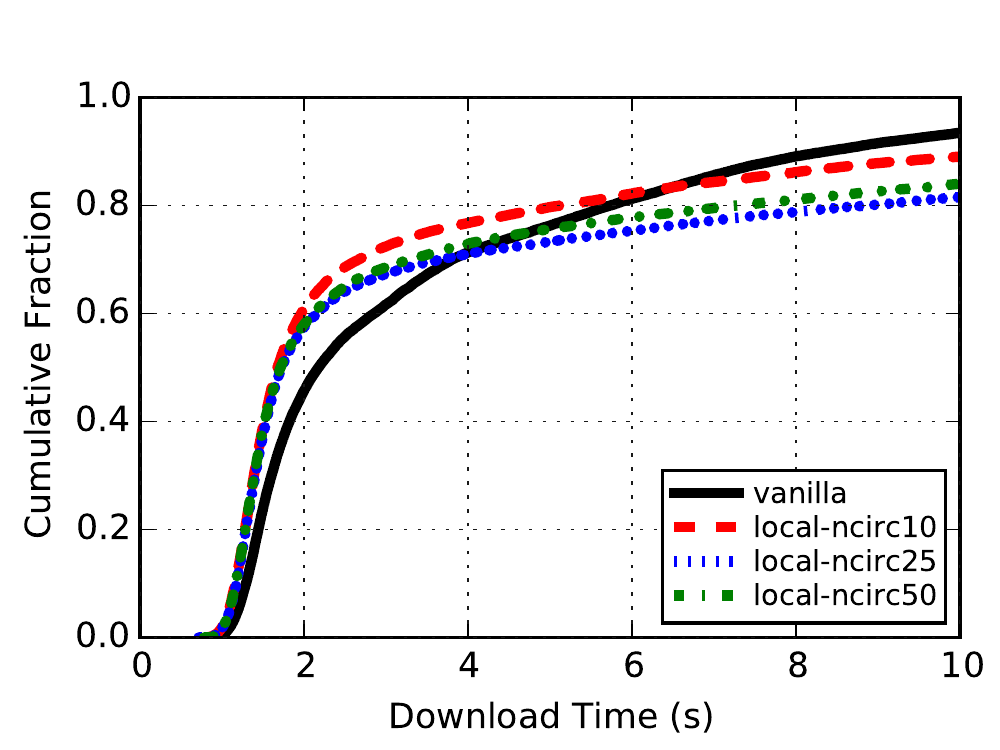}}
    }
    \subfloat[Bulk Download Times]
    {
        \label{fig:local-ncirc-ttlb-bulk}
        {\includegraphics[width=0.23\textwidth]{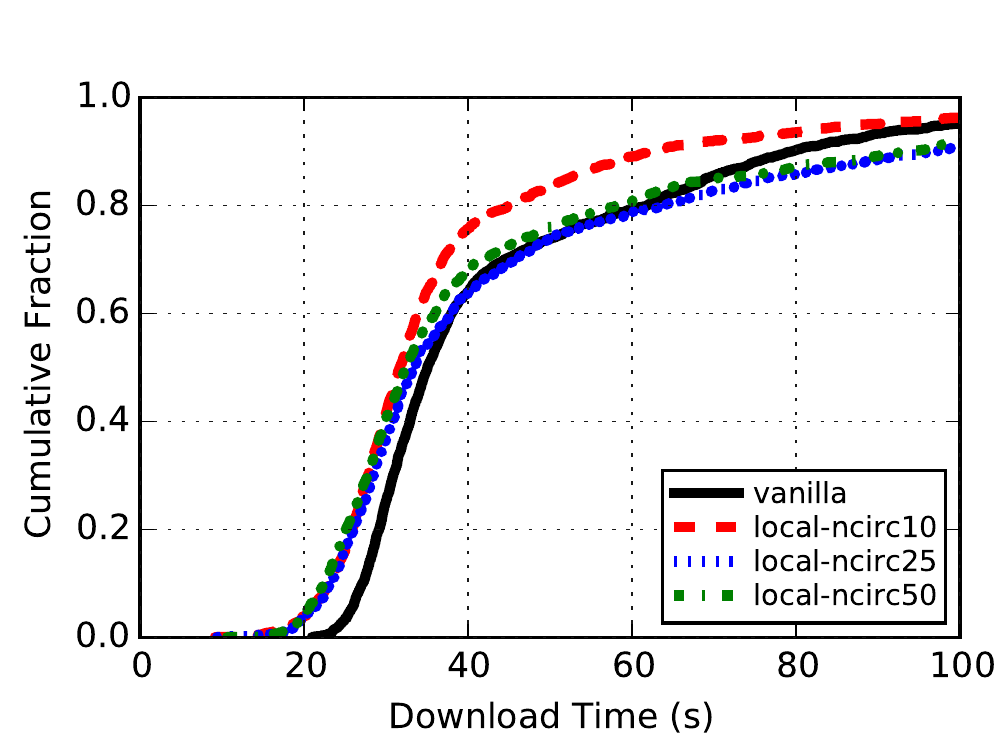}}
    }
    \subfloat[Total Client Bandwidth]
    {
        \label{fig:local-ncirc-client-bw}
        {\includegraphics[width=0.23\textwidth]{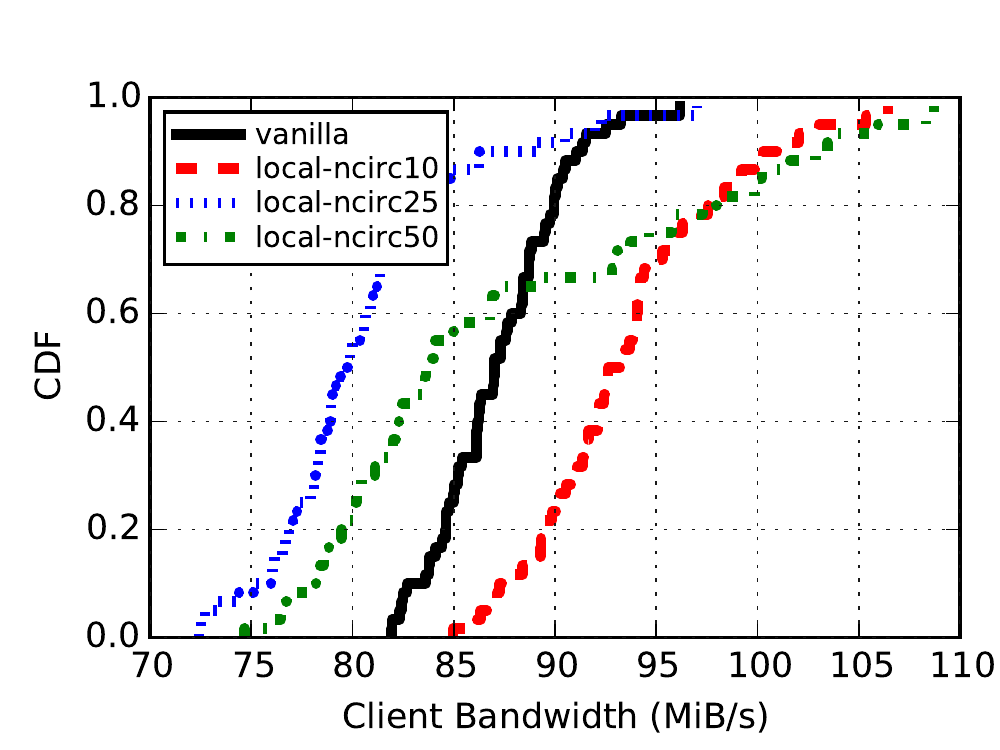}}
    }
    \caption
    {
        Download times and client bandwidth when using local decentralized
        circuit selection, varying the number of available circuits to select
        from.
    }
    \label{fig:local-ncirc-results}
\end{figure*}

\begin{figure*}[t]
    \centering
    \subfloat[Time to First Byte]
    {
        \label{fig:central-ncirc-ttfb}
        {\includegraphics[width=0.23\textwidth]{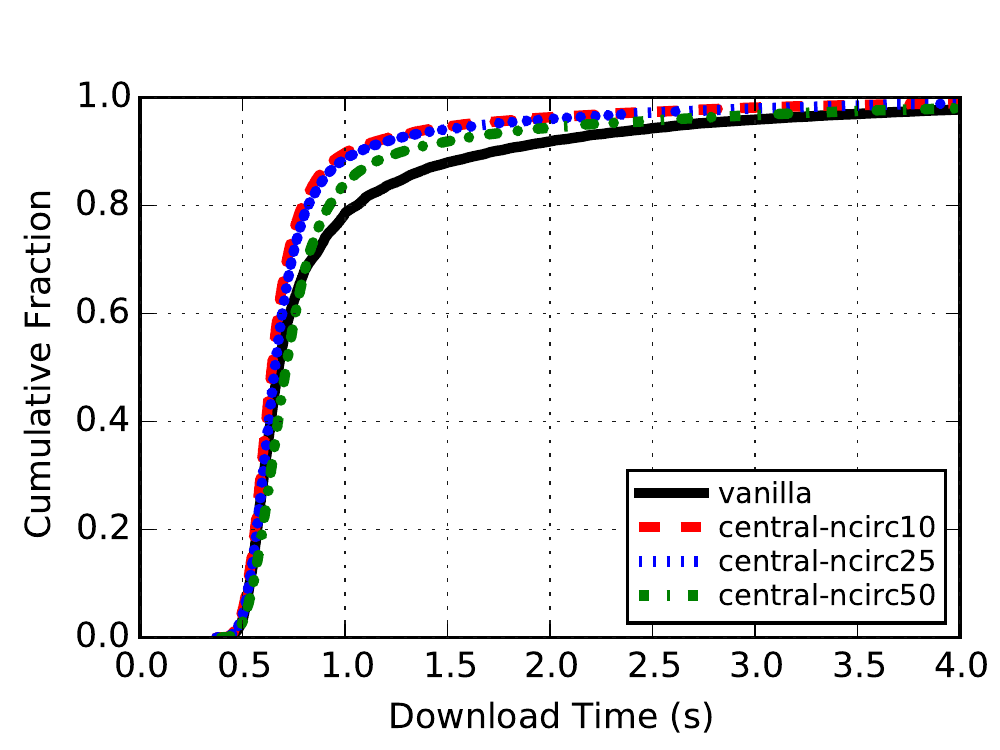}}
    }
    \subfloat[Web Download Times]
    {
        \label{fig:central-ncirc-ttlb-web}
        {\includegraphics[width=0.23\textwidth]{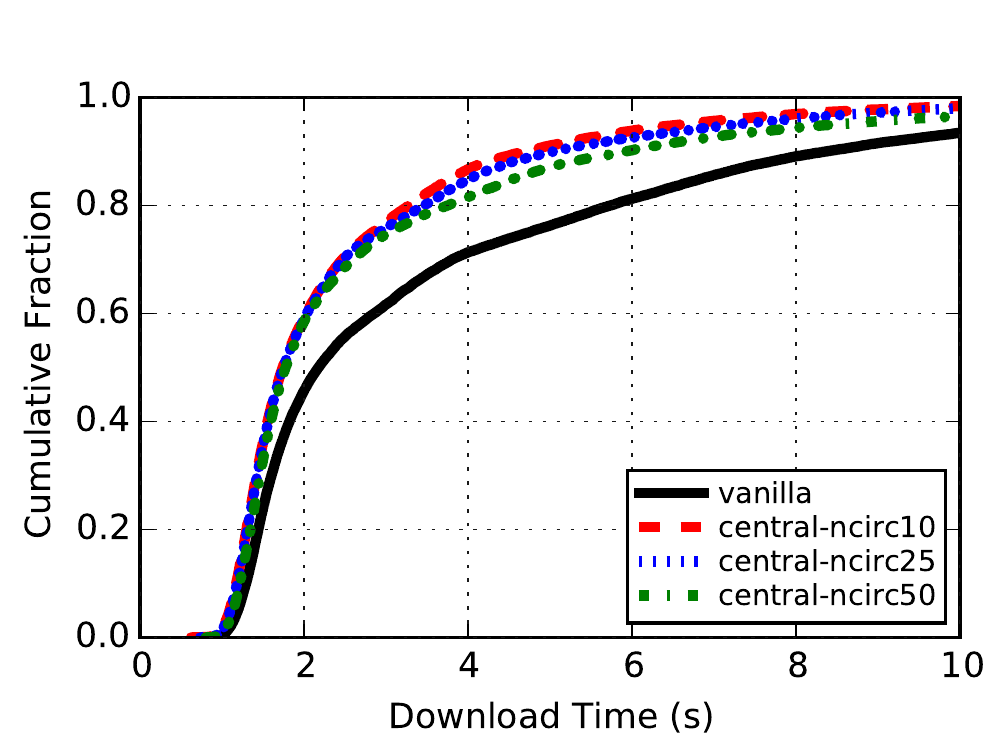}}
    }
    \subfloat[Bulk Download Times]
    {
        \label{fig:central-ncirc-ttlb-bulk}
        {\includegraphics[width=0.23\textwidth]{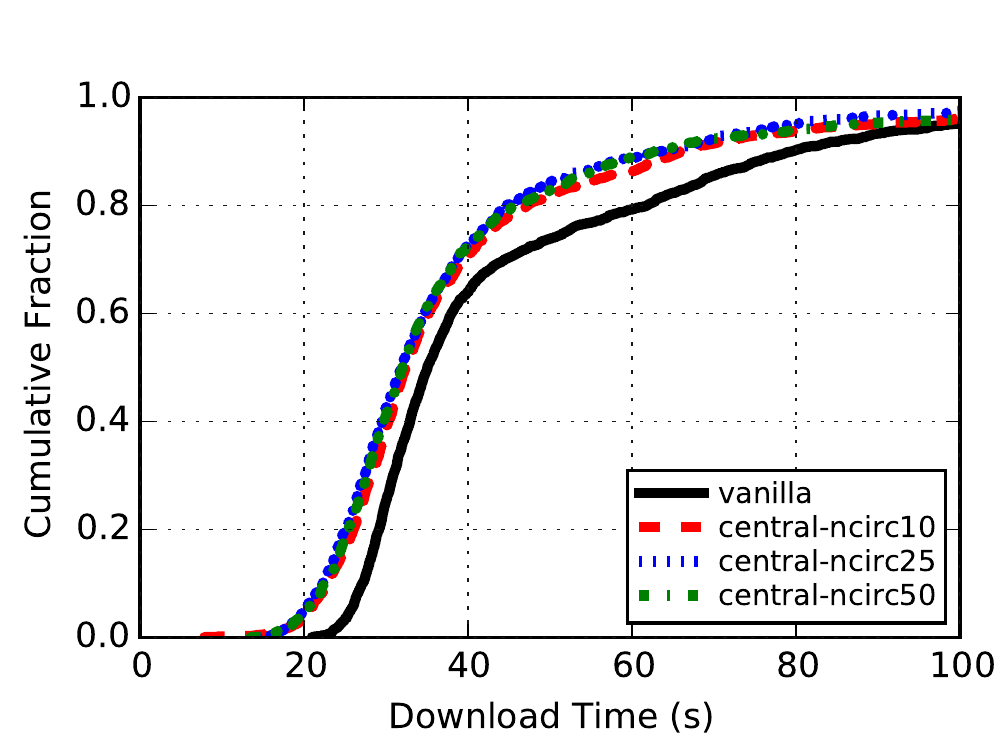}}
    }
    \subfloat[Total Client Bandwidth]
    {
        \label{fig:central-ncirc-client-bw}
        {\includegraphics[width=0.23\textwidth]{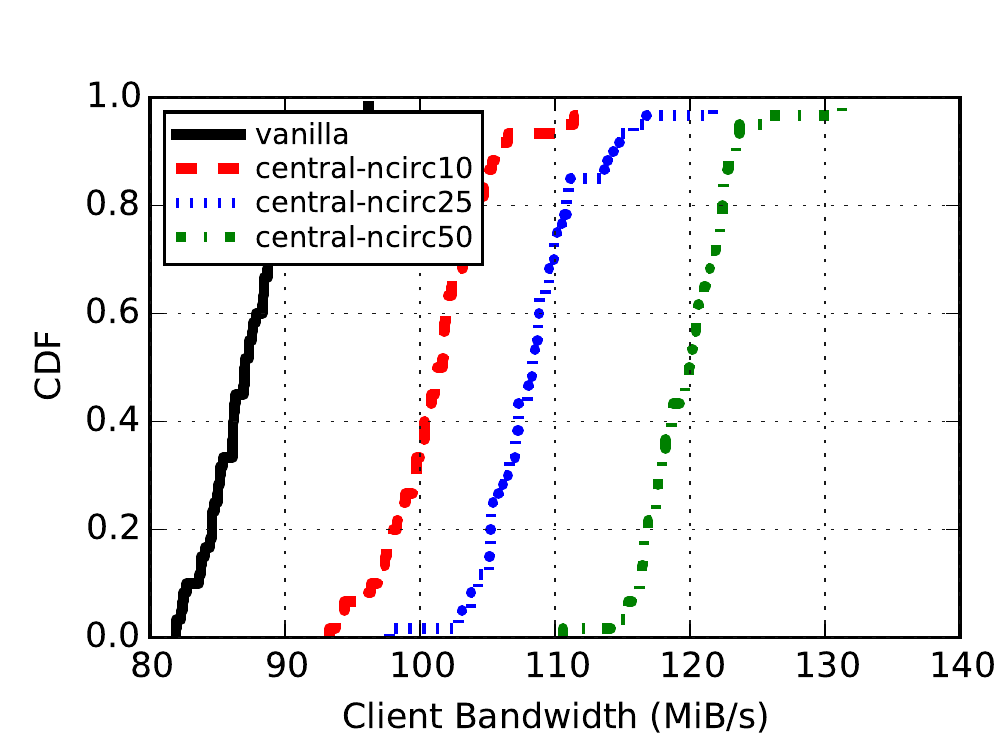}}
    }
    \caption
    {
        Download times and client bandwidth when using centralized circuit
        selection, varying the number of available circuits to select from.
    }
    \label{fig:central-ncirc-results}
\end{figure*}

Using these parameters we ran an experiment with the decentralized DWC
algorithm, along with vanilla Tor and centralized DWC.  Note that
\textit{every} experiment was configured to have relays send gossip cells every
5 seconds in addition with including a central authority connected to every
client.  In non-centralized experiments, when a download starts the central
authority tells the client to select a circuit themselves instead of making the
selection for them, and we had vanilla Tor ignore relay weights when making
circuit selection.  This is done to ensure that any latency and bandwidth
overhead is identical across experiments and the only difference is the circuit
selection algorithm being used.  The results of all three experiments are shown
in Figure~\ref{fig:local-central-results}.  The centralized experiments show
across the board improvements in every metric compared to vanilla Tor, with
clients experiencing almost 20\% higher bandwidth.  While the decentralized
experiment produced results slightly under the centralized, we still see
improvements compared to vanilla Tor.
Figure~\ref{fig:local-central-client-bw} shows client bandwidth still
increasing 8\% with the decentralized algorithm.  Bulk clients in both the
centralized and decentralized experiments achieved between 5\% and 30\% faster
downloads. While all web clients saw faster download times in the centralized
experiment, the results from the decentralized experiment are slightly more
mixed.  70\% of web client downloads performed just as fast as they did in
the centralized experiment, with 15\% falling in between centralized and
vanilla Tor.  The final 15\% of web client downloads experienced just slightly
longer download times, with 1-2\% increases compared to vanilla Tor. Time to
first byte shown in Figure~\ref{fig:local-central-ttfb} shows the worst
performance in the decentralized experiments, with roughly 25\% of downloads
experiencing an extra 1-2 seconds to receive their first byte.  This could be
caused by the fact that there is a delay in a relay updating its weight and all
clients receiving it, causing clients to select relays that are temporarily
congested resulting in poor times to first byte.

Along with the discussed parameters for the local weight computation, both the
centralized and decentralized have an \textit{implicit} parameter, the number
of active circuits client should attempt to maintain.  By default Tor clients
keep around 10 active circuits it can select from.  This is to make sure that
there is always some circuit that can be used for downloads, and the client
doesn't have to incur the 5-10 second overhead of circuit creation when a
download starts.  For both the centralized and decentralized algorithms, having
more circuits available to select from could increase performance.
As was seen in \S{ssec:online} when we had the full set of circuits to select
from, the centralized DWC algorithm produced the best results.  To test this
we configured the experiments to have clients maintain a set of 10, 25, and 50
available circuits.  Results for these experiments are shown in
Figures~\ref{fig:local-ncirc-results} and \ref{fig:central-ncirc-results}.
Across almost every metric the decentralized algorithm saw \textit{worse}
results when using more circuits.  This is particularly evident looking at
client bandwidth in Figure~\ref{fig:local-ncirc-client-bw}. When selecting from
50 circuits only about 40\% of the time clients achieved a higher bandwidth,
and using 25 circuits \textit{always} resulted in worse performing clients.
This is most likely due to the fact that increasing the number of circuits that
clients create is interfering with the bottleneck classification algorithm,
resulting in worse performance. For the centralized algorithm 
all download times remained fairly identical no matter how many circuits were
available.  The only metric that saw any difference was client bandwidth shown
in Figure~\ref{fig:central-ncirc-client-bw}.  Here we see increasing the number
of available circuits to 25 and 50 achieves an increase of 8\% and 19\% in client
bandwidth compared to using only 10 circuits. Considering that no other metrics
improved when increasing this is most likely due to overhead from maintaining
circuits that is not being controlled for.

\begin{figure*}[t!]
    \centering
    \subfloat[]
    {
        \label{fig:local-privacy-correlation}
        {\includegraphics[width=0.32\textwidth]{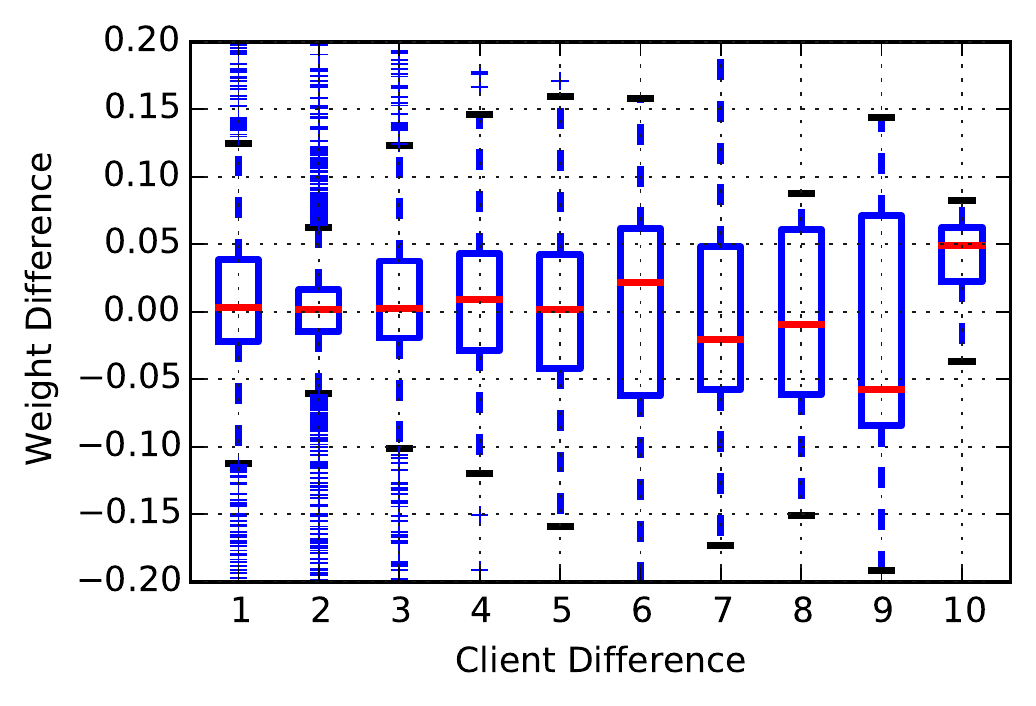}}
    }
    \subfloat[]
    {
        \label{fig:usage-cdf}
        {\includegraphics[width=0.32\textwidth]{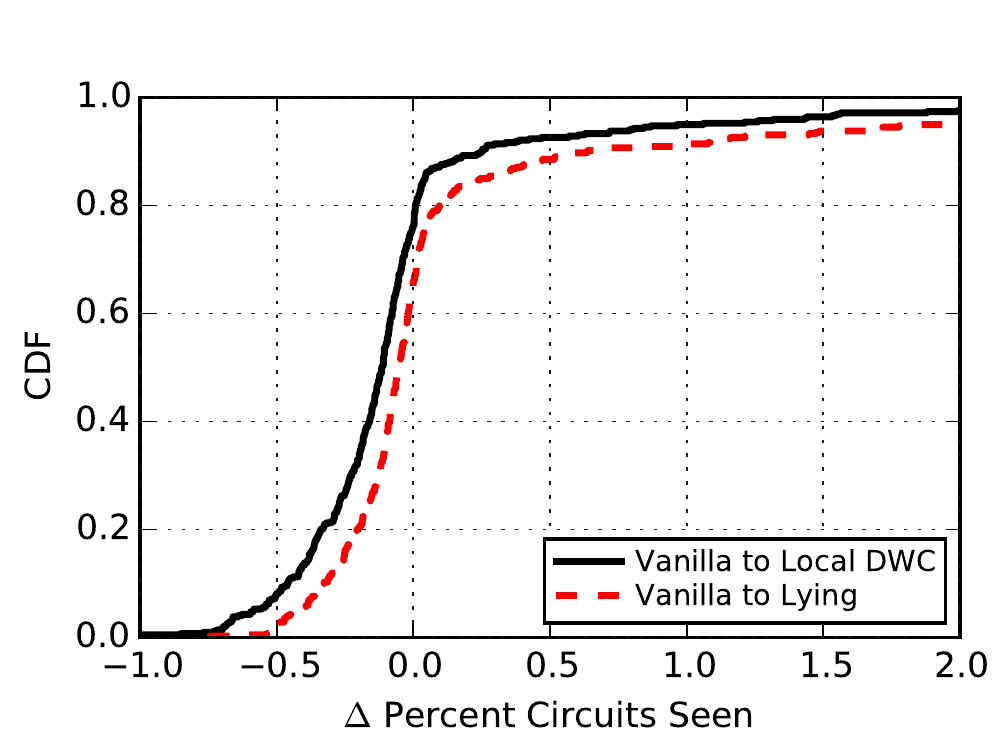}}
    }
    \subfloat[]
    {
        \label{fig:usage-top}
        {\includegraphics[width=0.32\textwidth]{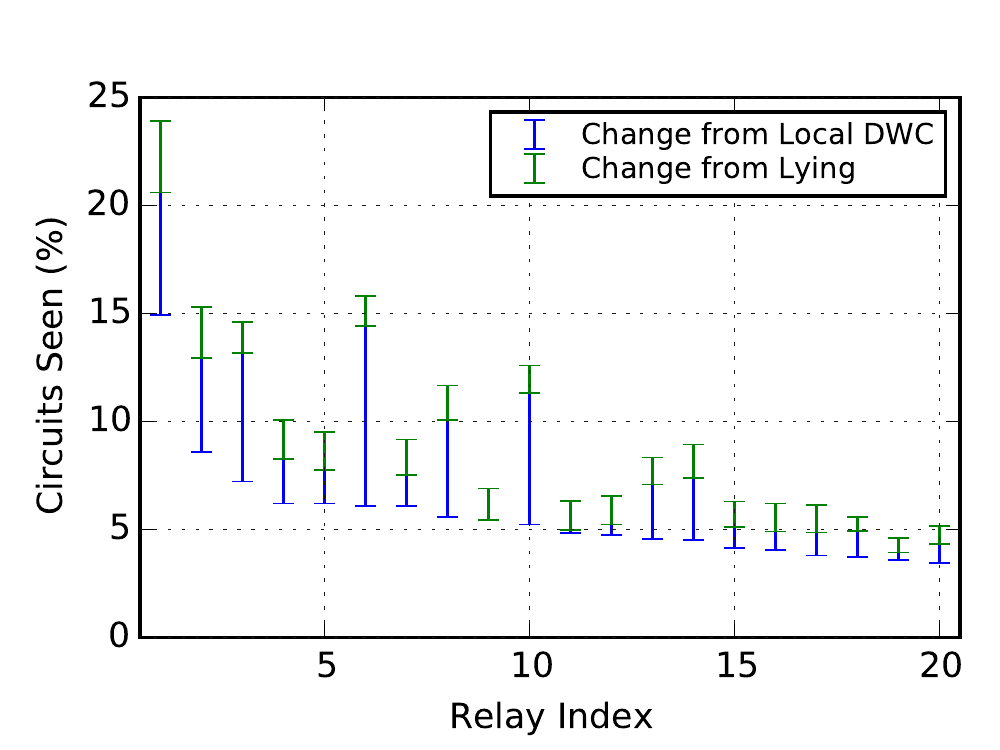}}
    }
    \caption
    {
        (a) weight difference on relays compared to the difference in number of
        bottleneck circuits on the relay 
        (b) CDF of the difference in percent of circuits a relay sees when a
        relay lies about their weight 
        (c) percent of circuits a relay is on for the top 20 relays going from
        vanilla Tor to decentralized DWC (blue), in addition with the gains achieved
        when relays lie about their weight (green)
    }
    \label{fig:lying}
\end{figure*}

\subsection{Privacy Analysis} \label{ssec:privacy}

With the addition of gossip cells and the new decentralized circuit selection
protocol, there are some avenues that could be abused in an attempt to reduce
client anonymity.  The first major concern is that since the gossip cells
purposefully leak information about the internal state of the relays, that this
could open up a new side channel where an adversary can correlate client
activity to change in information leaked to determine which relays are used by
a client.  The second issue is the increase in percent of circuits that relays
end up being selected on, both from the change in selection algorithm and also
due to the fact that an adversarial relay could abuse the protocol.  This can
be done by lying about their own weight or artificially inflating other relays
weight, thereby increasing the chance they will be selected allowing them to
observe a higher portion of the Tor network.  In this section we examine the
impact of some of these privacy issues.

\begin{figure*}[t!]
    \centering
    \subfloat[]{\label{fig:dos-weight}{\includegraphics[width=0.32\textwidth]{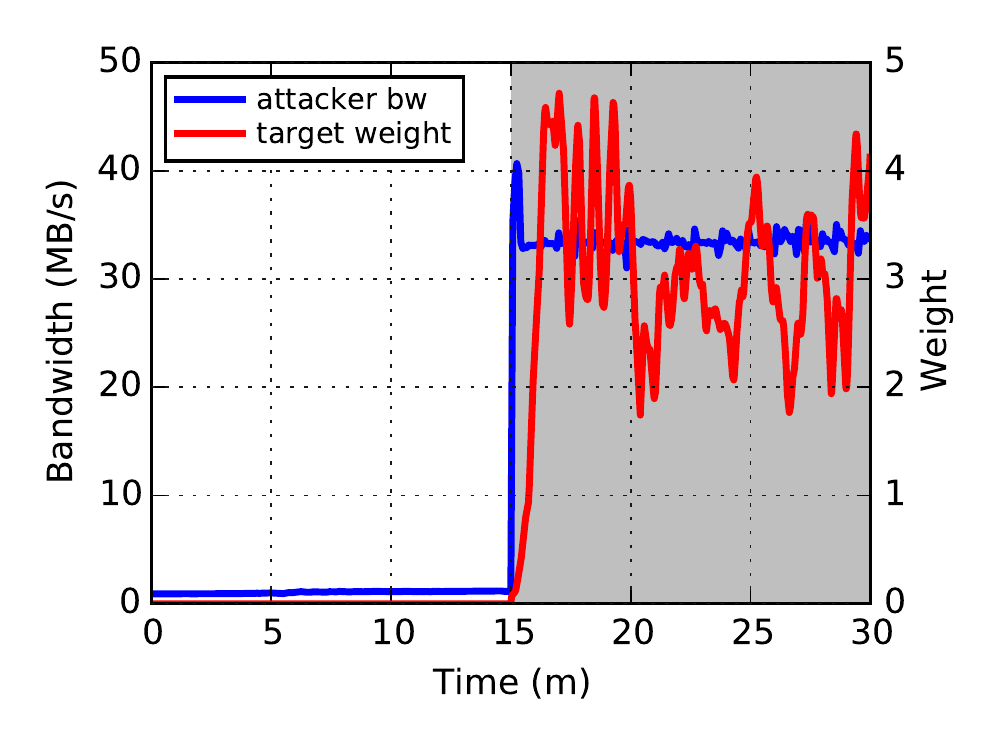}}}
    \subfloat[]{\label{fig:dos-circuits}{\includegraphics[width=0.32\textwidth]{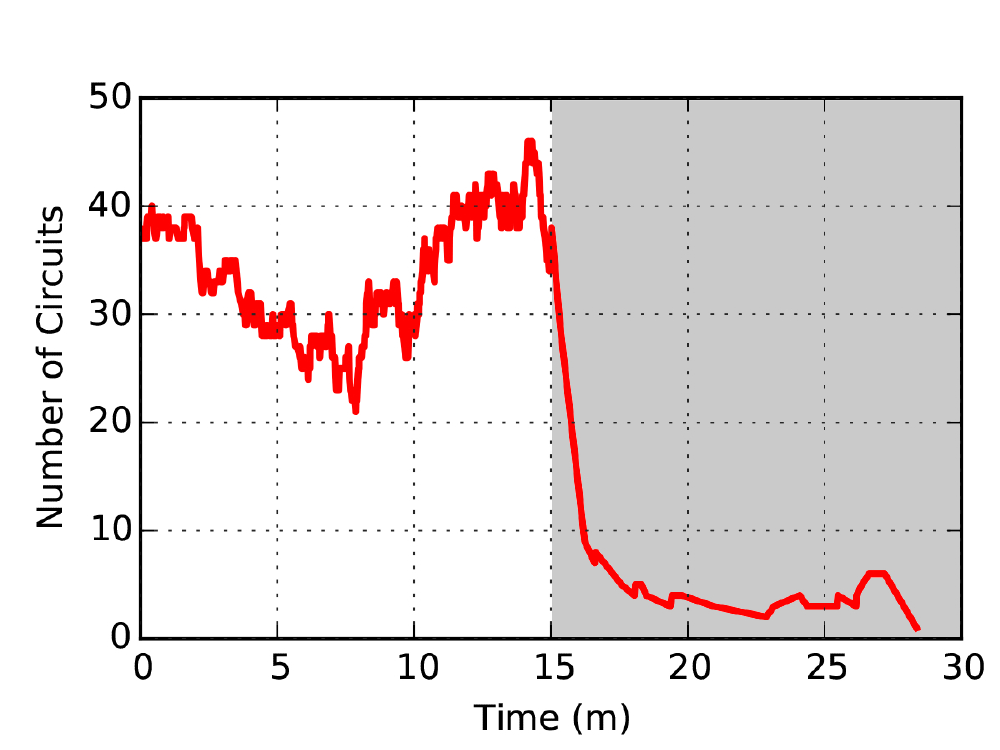}}}
    \subfloat[]{\label{fig:dos-finished}{\includegraphics[width=0.32\textwidth]{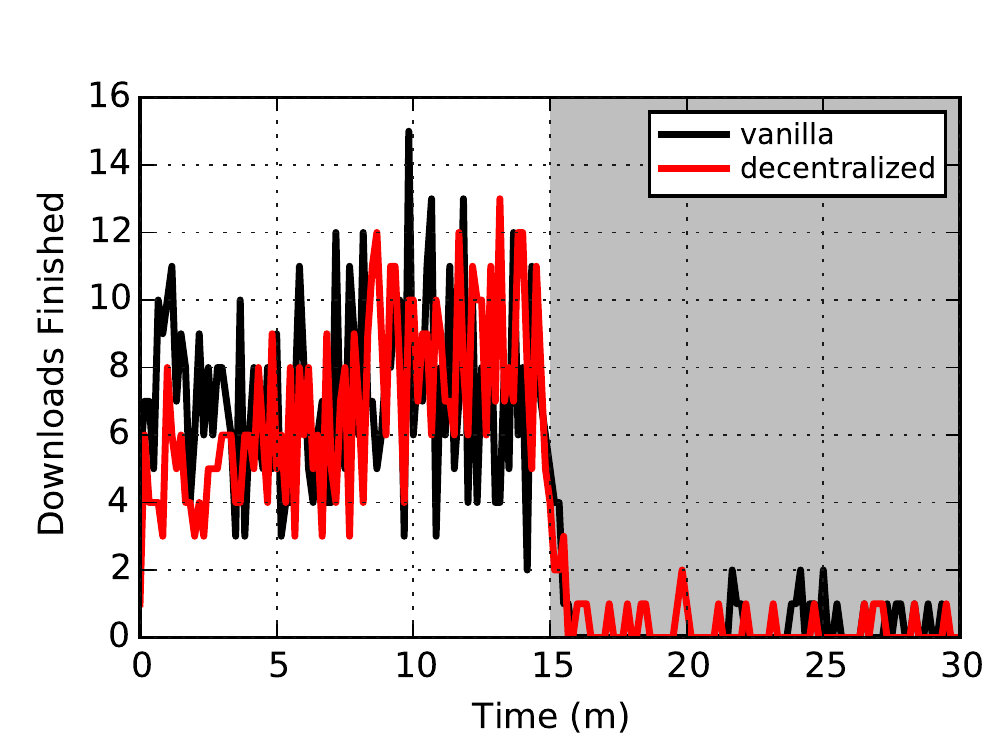}}}
    \caption{(a) attackers bandwidth and targets weight when an adversary is
        running the denial of service attack (b) number of clients using the
        target for a circuit (c) number of downloads completed by clients using
        the target relay before and after attack is started, shown both for
    vanilla Tor and using decentralized circuit selection}
    \label{fig:dos-attack}
\end{figure*}

\subsubsection{Information Leakage}

The first issue is that relays advertising their local weight calculation
could leak some information to an adversary.  Mittal \etal
\cite{mittal2011stealthy} showed how an adversary could use throughput
measurements to identify bottleneck relays in circuits.  Relay weight values
could be used similarly, where an adversary could attempt to correlate start
and stop times of connections with the weight values of potential bottleneck
relays.  To examine how much information is leaked by the weight values, every
time a relay sent a \verb|GOSSIP| cell we recorded the weight of the relay and
the number of active circuits using the relay.  Then for every two consecutive
\verb|GOSSIP| cells sent by a relay we then recorded the difference in weight
and number of active circuits, which should give us a good idea of how well
the change in these two values are correlated over a short period of time.
Figure~\ref{fig:local-privacy-correlation} shows the distribution of weight
differences across various changes in clients actively using the relay.  Since
we are interested in times when the relay is a bottleneck on new circuits, we
excluded times when the weight difference was 0 as this is indicative that the
relay was \textit{not} a bottleneck on any of the new circuits.  This shows an
almost nonexistent correlation between the two metrics, where we are just as
likely to see a rise \textit{or} drop in weight after more bottleneck circuits
start using a relay.  In the situation similar to the one outlined in
\cite{mittal2011stealthy} where an adversary is attempting to identify
bottleneck relays used in a circuit, we are particularly interested in the
situation where the number of active circuits using the relay as a bottleneck
increases by 1.  If there were large (maybe temporary) changes noticeable to an
adversary they could identify the bottleneck relay.  But as we can see in
Figure~\ref{fig:local-privacy-correlation} the distribution of weight changes
when client difference is 1 is very large, meaning it would be almost
impossible to identify the bottleneck relay by just monitoring relay weights.

\subsubsection{Relay Usage Changes}

Since relays are still selected at random weighted by their bandwidth, over the
long term relay usage should remain the same in vanilla Tor and when using
the decentralized DWC circuit selection algorithm. Over the short term this
might change though, since circuits are selected based on network information
that could cause peeks in relay usage.  In addition to short term fluctuations,
an active adversarial relay could \textit{lie} about their DWC weight,
consistently telling clients they have a weight of 0, increasing the chances
clients select circuits that the adversaries relay lies on.  These changes
could allow a relay to increase their view of the network without having to
actually provide more bandwidth, potentially reducing anonymity of clients
using Tor.

To determine the effect that the decentralized algorithm has on relay usage,
we extracted all circuits selected by clients during each experiment.  In
addition, for the experiment running the decentralized circuit select algorithm
we had clients output every circuit it considered, along with each relays
weight.  This allows us to perform a static analysis of which circuits
\textit{would have} been selected if a single relay was lying about their
weight.  In each situation we are interested in the change in the percent of
circuits selected containing a specific relay, showing how much more of the
network an adversary might be able to see. Figure~\ref{fig:usage-cdf} shows the
CDF of the change in percent of circuits experienced by each relay. It shows
the difference when going from vanilla Tor to regular decentralized DWC, in
addition to going from vanilla Tor to decentralized DWC when relays lie about
their weight.  While the increases are generally small, the largest changes
come from simply using the new decentralized algorithm, with 80\% of relays
seeing a small \textit{drop} in the percent of circuits selected they end up
on.  Even the relays that see an increase almost never end up seeing more than 1\%
additional circuits.  Furthermore there is very little gained when relays start
lying about their weight, with most increases being an extra 0.1-0.2\%.
Figure~\ref{fig:usage-top} shows the 20 relays with the largest gains.  These
are the highest bandwidth relays in the network, providing 39\% of the total
available bandwidth.  Here we see in the most extreme case, the very highest
bandwidth relay could view 9\% more circuits than they would have seen in
vanilla Tor over the short term.  Over the long term however, the only real
gain coming from the relay lying about their weight would be capped at about
3-4\%.

\subsubsection{Denial of Service}

While adversarial relays are limited in the number of extra circuits they can
be placed on by lying about their weight, they still could have the ability to
reduce the chances that other relays are selected.  To achieve this they would
need to artificially inflate the weight of other relays in the network,
preventing clients from selecting circuits that the target relays appear on.
Recall that the local weight calculation is based on how many
\textit{bottleneck} circuits the relay estimates they are on.  This means that
an adversary cannot simply just create inactive circuits through the relay to
inflate their weight, those circuits would never be labeled as bottleneck.  So
to actually cause the weight to increase the adversary needs to actually send
data through the circuits.  To test the effectiveness, we configured an
experiment to create 250 one-hop circuits through the target relay.  After 15
minutes the one-hop circuits were activated, downloading as much data as they
could.  Note that we want as many circuits through the relay as possible to
make their weight as large as possible.  The relay weight is summed across all
bottleneck circuits, $\sum bw(c_i)^{-1}$.  If we have $n$ one-hop circuits
through a relay of bandwidth $bw$, each circuit will have a bandwidth of
roughly $\frac{bw}{n}$, so the weight on the relay will be $\sum_1^n
\left(\frac{bw}{n}\right)^{-1} = \frac{n^2}{bw}$. 

Figure~\ref{fig:dos-weight} looks at the weight of the target relay along with
how much bandwidth the attacker is using, with the shaded region noting when
the attack is active.  We see that the attacker is able to push through close
to the maximum bandwidth that the relay can handle, around 35 MB/s. When the
circuits are active the weight spikes to almost 100 times what it was
previously.  Figure~\ref{fig:dos-circuits} shows the number of circuits the
target is actively on.  After the attack is started the number plummets to
almost 0, down from the 30-40 it was previously on.  But while the attack does
succeed in inflating the weight of the target, note that the adversary has to
fully saturate the bandwidth of the target.  Doing this in vanilla Tor will
have has almost the same effect, essentially running a denial of service by
consuming all the available bandwidth.  Figure~\ref{fig:dos-finished} looks at
the number of completed downloads using the target relay in both vanilla Tor
and when using the decentralized circuit selection.  Both experiments see the
number of successful downloads drop to 0 while the attack is running.  So even
though the addition of the relay weight adds another mechanism that can be used
to run a denial of service attack, the avenue (saturating bandwidth) is the
same.

\section{Conclusion} \label{sec:conclusion}

In this paper we explore the price of anarchy in Tor by examining how using
offline and online algorithms for circuit selection is able to significantly
increase performance to end clients.  These algorithms are able to more
efficiently utilize network resources, resulting in almost twice as much
bandwidth being consumed.  Furthermore, since the offline and online produced
almost identical results, it suggests that that network utilization was near
or at capacity.  In addition we adapt the online DWC algorithm to create a
decentralized version, where relays themselves are able to estimate
calculations made by a central authority.  While the results are not quite as
dramatic, the decentralized algorithm was still able to achieve 20-25\% higher
bandwidth consumption when compared to vanilla Tor.  Finally we analyze the
potential loss in anonymity when using the decentralized algorithm, both
against a passive and active adversary.

\section*{Acknowledgments}
This research was supported by NSF grant 1314637.

\bibliographystyle{abbrv}
\bibliography{ref}


\appendix

\section{Bandwidth Algorithm Proof} \label{sec:bwalgproof}

Let $R$ be the relay selected with $B$ bandwidth and $C$ circuits.  Let $R'$ be
a different relay with $B'$ bandwidth and $C'$ circuits.  By definition $R$ is
selected such that $\frac{B}{C} \leq \frac{B'}{C'}$.  When iterating through
the $C$ circuits let $n$ be the number that $R'$ is on.  Note that this means
that $n \leq C'$.  After the circuits have been iterated through and operations
performed, $R'$ will have $B' - \frac{B}{C} \cdot n$ bandwidth left with $C' -
n$ circuits.

Assume that $R'$ has 0 bandwidth afterwards, so $B' - \frac{B}{C} \cdot n = 0$.  We
want to show this means that $R'$ is on no more circuits so that $C' - n = 0$.
We have
\[
    B' - \frac{B}{C} \cdot n = 0 \Rightarrow B' = \frac{B}{C} \cdot n
    \Rightarrow n = \frac{B' \cdot C}{B}
\]
So that means that the number of circuits $R'$ is left on is
\begin{eqnarray*}
    C' - n &=& C' - C - \frac{B' \cdot C}{B} = \frac{B \cdot C'}{B} - \frac{B'
    \cdot C}{B} \\
    &=& \frac{B \cdot C' - B' \cdot C}{B}
\end{eqnarray*}
However, $R$ was picked such that 
\[
    \frac{B}{C} \leq \frac{B'}{C'} \Rightarrow B \cdot C' \leq B' \cdot C
    \Rightarrow B \cdot C' - B' \cdot C \leq 0
\]
This gives us
\[
    C' - n = \frac{B \cdot C' - B' \cdot C}{B} \leq 0
\]
since we know $B > 0$ and
\[
    n \leq C' \Rightarrow 0 \leq C' - n
\]
which implies that $0 \leq C' - n \leq 0 \Rightarrow C' - n = 0$.

\section{Circuit Pruning Algorithm} \label{sec:prunealg}

\begin{algorithm}[h]
    \caption{Generate pruned circuit set}
    \label{alg:genpruned}
    \begin{algorithmic}[1]
        \Function{BuildPrunedSet}{$relays$}
            \State $circuits \gets List()$
            \While{$TRUE$}
                \If{$relays.len() > 3$}
                    \State \bf{break}
                \EndIf
            
                \If{$relays.numExits() > 0$}
                    \State \bf{break}
                \EndIf

                \State $relays.sortByBW()$
                \State $exit \gets relays.getFirstExit()$
                \State $middle \gets relays.getFirstNonExit()$
                \State $guard \gets relays.getFirstNonExit()$

                \If{$middle == Null$}
                    \State $middle \gets relays.getFirstExit()$
                \EndIf
                \If{$guard == Null$}
                    \State $guard \gets relays.getFirstExit()$
                \EndIf

                \State $circuits.append(guard,middle,exit)$

                \State $bw \gets min(guard.bw, middle.bw, exitbw)$
                \State $guard.bw \gets guard.bw - bw$
                \State $middle.bw \gets middle.bw - bw$
                \State $exit.bw \gets exit.bw - bw$

                \If{$guard.bw == 0$}
                    \State $relays.remove(guard)$
                \EndIf
                \If{$middle.bw == 0$}
                    \State $relays.remove(middle)$
                \EndIf
                \If{$exit.bw == 0$}
                    \State $relays.remove(exit)$
                \EndIf
            \EndWhile

            \State \bf{return }$circuits$
                
        \EndFunction
    \end{algorithmic}
\end{algorithm}

\end{document}